\documentclass[]{aastex631}

\usepackage{amsmath}
\usepackage{amssymb}
\usepackage{tikz}
\usepackage{color}

\newcommand{\blue}{\textcolor{black}}

\usetikzlibrary{calc,shapes.symbols,decorations.pathreplacing,shapes.geometric}
\tikzstyle{process1} = [ellipse, text centered, draw=black, fill=red!30]
\tikzstyle{process2} = [ellipse, text centered, draw=black, fill=blue!30]

\allowdisplaybreaks

\begin{document}

\title{Kinetic equilibrium of two-dimensional force-free current sheets}

\correspondingauthor{Xin An}
\email{phyax@ucla.edu}

\author[0000-0003-2507-8632]{Xin An}
\affiliation{Department of Earth, Planetary, and Space Sciences, University of California, Los Angeles, CA, 90095, USA}

\author[0000-0001-8823-4474]{Anton Artemyev}
\affiliation{Department of Earth, Planetary, and Space Sciences, University of California, Los Angeles, CA, 90095, USA}
\affiliation{Space Research Institute of the Russian Academy of Sciences, Moscow, 117997, Russia}


\author[0000-0001-7024-1561]{Vassilis Angelopoulos}
\affiliation{Department of Earth, Planetary, and Space Sciences, University of California, Los Angeles, CA, 90095, USA}

\author[0000-0001-5544-9911]{Andrei Runov}
\affiliation{Department of Earth, Planetary, and Space Sciences, University of California, Los Angeles, CA, 90095, USA}

\author[0000-0002-2261-0331]{Sergey Kamaletdinov}
\affiliation{Space Research Institute of the Russian Academy of Sciences, Moscow, 117997, Russia}
\affiliation{Faculty of Physics, National Research University Higher School of Economics, Moscow, 101000, Russia}



\begin{abstract}
Force-free current sheets are local plasma structures with field-aligned electric currents and approximately uniform plasma pressures. Such structures, widely found throughout the heliosphere, are sites for plasma instabilities and magnetic reconnection, the growth rate of which is controlled by the structure's current sheet configuration. Despite the fact that many kinetic equilibrium models have been developed for one-dimensional (1D) force-free current sheets, their two-dimensional (2D) counterparts, which have a magnetic field component normal to the current sheets, have not received sufficient attention to date. Here, using particle-in-cell simulations, we search for such 2D force-free current sheets through relaxation from an initial, magnetohydrodynamic equilibrium. Kinetic equilibria are established toward the end of our simulations, thus demonstrating the existence of kinetic force-free current sheets. Although the system currents in the late equilibrium state remain field aligned as in the initial configuration, the velocity distribution functions of both ions and electrons systematically evolve from their initial drifting Maxwellians to their final time-stationary Vlasov state. The existence of 2D force-free current sheets at kinetic equilibrium necessitates future work in discovering additional integrals of motion of the system, constructing the kinetic distribution functions, and eventually investigating their stability properties.
\end{abstract}



\section{Introduction} \label{sec:intro}
Current sheets are spatially localized plasma structures that play an essential role in various space plasma systems: solar flares \citep{Syrovatskii81,bookParker94,Fleishman&Pevtsov18},  solar wind turbulence \citep{Servidio11:npg,Borovsky10:solarwind,Vasko22:apjl}, boundaries of planetary magnetospheres – magnetopauses \citep{deKeyser05}, magnetotails of planets \citep{Jackman14,Achilleos18GMS:jupiter,Lui18} and comets \citep{Cravens&Gombosi04,Volwerk18A&A,Volwerk18GMS}. Strong currents flowing within current sheets are subject to various instabilities resulting in magnetic field line reconnection, which converts magnetic energy to plasma heating and particle acceleration \citep[e.g.,][]{book:Gonzalez&Parker,bookBirn&Priest07}. 

\begin{figure}[tphb]
    \centering
    \includegraphics[width=1\textwidth]{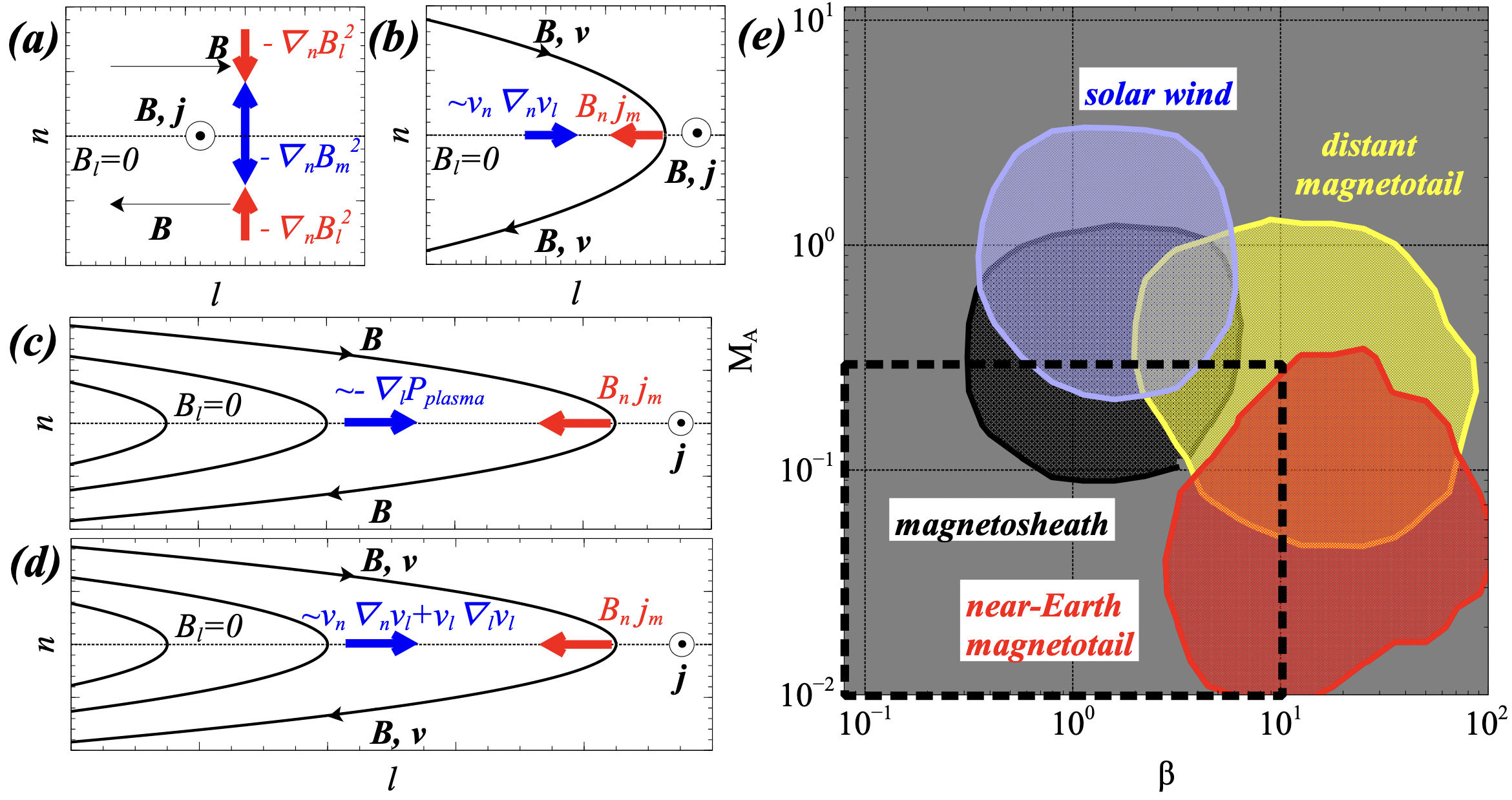}
    \caption{Panels (a)-(d) show magnetic field lines and the main components of stress balance for different current sheet configurations. The coordinate system consists of the $l$ component along the main magnetic field direction, reversing sign at the current sheet neutral plane ($B_l=0$), the $m$ component along the main current density direction corresponding to variations of $B_l$, and the $n$ (normal) component along the spatial gradient of $B_l$ (i.e., $4\pi j_m/c=\partial B_l/\partial r_n$).  Panel (e) shows the typical parameter regimes of current sheets observed by THEMIS and ARTEMIS \citep{Angelopoulos08:ssr,Angelopoulos11:ARTEMIS} in the solar wind (blue), Earth's magnetosheath (shocked solar wind; black), near-Earth magnetotail (red), and lunar-distance magnetotail (green). The black dashed box defines the parameter regime where 2D force-free current sheets are expected. Our dataset includes $\sim300$ solar wind current sheets, $\sim100$ magnetosheath current sheets, $\sim 100$ current sheets in the near-Earth magnetotail, and $\sim 500$ current sheets in the lunar-distance magnetotail. All current sheets were identified from THEMIS and ARTEMIS fluxgate magnetometer measurements \citep{Auster08:THEMIS}. The selection procedure and data processing for solar wind current sheets are described in \citet{Artemyev19:jgr:solarwind}: ion temperatures are calculated using the OMNI dataset \citep{King&Papitashvili05,Artemyev18:jgr:report}, plasma flows in the current sheet reference frame are calculated as the change of the most variable flow component across the current sheet \citep[see details in][]{Artemyev19:jgr:solarwind}. The same criteria and data processing methods were applied to the magnetosheath (dayside) current sheets, but without using the OMNI data because THEMIS accurately measures magnetosheath ion temperatures in that region \citep{McFadden08:THEMIS}. The dataset of the near-Earth magnetotail current sheets (radial distances $\sim 10-30$ Earth radii) is described in \citet{Artemyev16:jgr:pressure}. The same criteria and data processing methods were applied to the lunar-distance current sheets. Details of the calculation of $\beta$ and $M_A$ are described in \citet{Artemyev19:jgr:solarwind,Artemyev17:jgr:mars}.  }
    \label{fig:scheme}
\end{figure}

The properties of magnetic reconnection and the dissipation rate of magnetic energy strongly depend on the current sheet configurations. The simplest magnetic field geometry is the one-dimensional (1D) current sheet, which is either a tangential discontinuity separating two plasmas with different properties or a rotational discontinuity having a finite magnetic field component, $B_n$, normal to the current sheet. The stress balance in 1D tangential discontinuities is established by the gradients of plasma and magnetic field pressures [see Figure \ref{fig:scheme}(a) and \citet{Allanson15,Neukirch20,Neukirch20:jpp}], whereas the stress balance in 1D rotational discontinuities requires a contribution from the plasma dynamic pressure in order to balance the magnetic field line tension force $\propto B_n$ [see Figure \ref{fig:scheme}(b) and \citet{Hudson70}].

\blue{In two-dimensional (2D) current sheets, $B_n\ne 0$ naturally appears \citep{bookSchindler06}; thus, the stress balance requires 2D plasma pressure gradients [see Figure \ref{fig:scheme}(c) and \citet{YL05}], 2D dynamic pressure gradients [see Figure \ref{fig:scheme}(d) and \citet{Birn92,Nickeler&Wiegelmann10,Cicogna&Pegoraro15}], or pressure anisotropy \citep[e.g.][]{Sitnov&Merkin16,Artemyev16:pop:cs}. The latter solution has multiple variants, most of which are devoted to the construction of quasi-1D current sheets with $B_n\ne 0$ and $\nabla_l \approx 0$. Such current sheet models aim to describe equilibria of planetary magnetospheres \citep[see discussion in][]{Sitnov06,Sitnov&Arnold22,Zelenyi11PPR,Zelenyi22} that often demonstrate a need for anisotropy contribution in the stress balance \citep[see discussion in][]{Sitnov19:jgr,Artemyev21:grl}. Moreover, nonlinear dynamics of large-amplitude Alfven waves in the solar wind and the subsequent formation of rotational discontinuities \citep[e.g.,][]{Medvedev96:pop,Medvedev97:pop,Vasquez&Hollweg98,Vasquez&Hollweg99} may also be affected by plasma anisotropy \citep[e.g.,][]{Tenerani17,Tenerani18}, with observable effects on discontinuity configuration \citep[see, e.g., discussion in][and references therein]{Artemyev20:apjl}. Thus plasma anisotropy provides an essential broadening of possible equilibrium configurations for current sheets with $B_n\ne0$. However, it is a less stable plasma parameter in the solar wind and magnetospheres, with almost no reliable statistics of anisotropy distribution around current sheets \citep[partially due to spacecarft instrumental limitations, see discussion in][]{Artemyev19:jgr:ions,Wilson22:frontiers}.} 

\begin{figure}[tphb]
    \centering
    \includegraphics[width=0.9\textwidth]{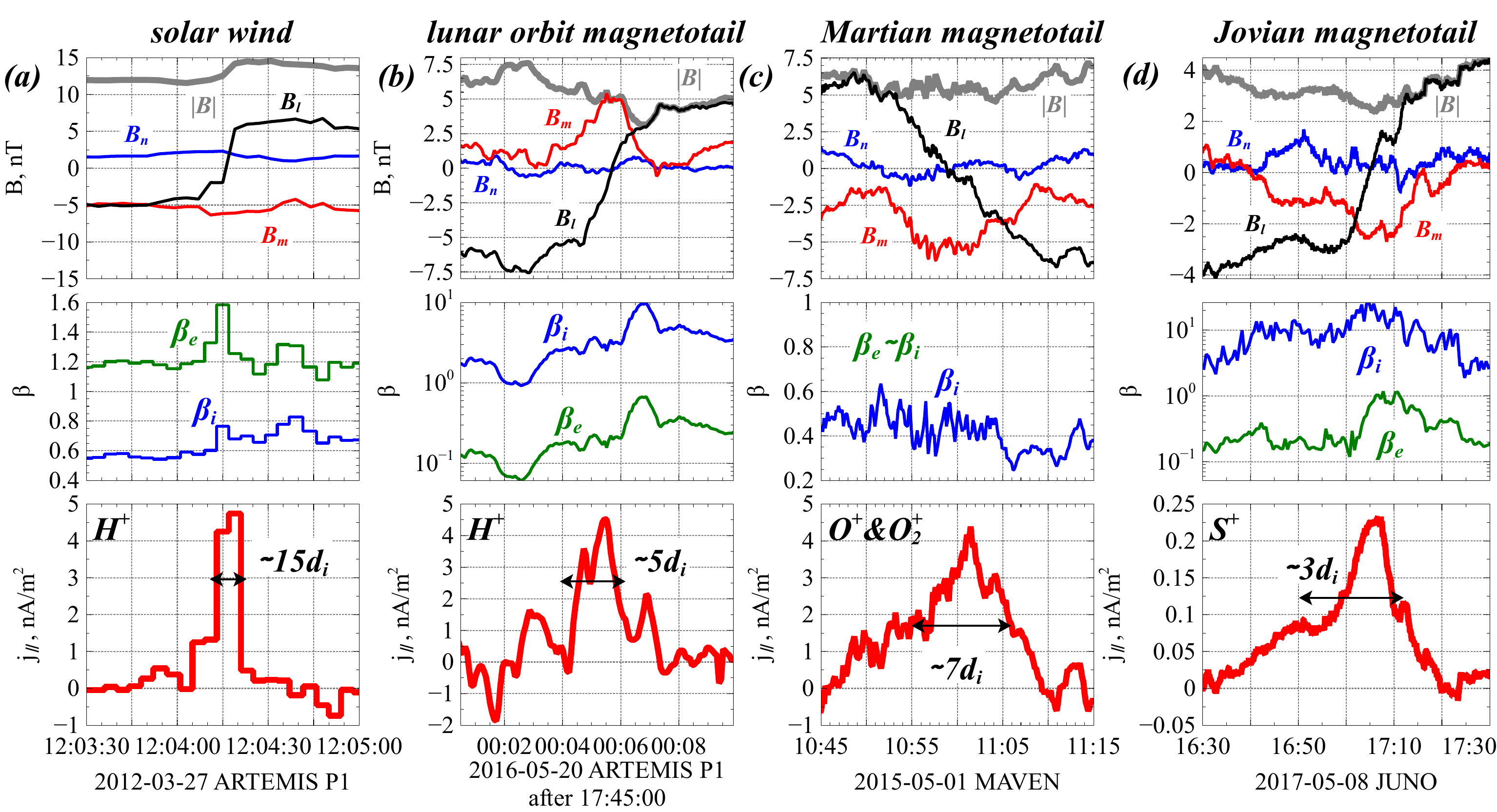}
    \caption{Four examples of force-free current sheets in (a) solar wind, (b) lunar-distance magnetotail, (c) Martian magnetotail, and (d) Jovian magnetotail. The top panels show the magnetic field in the local coordinate system \cite{Sonnerup68}. Grey curves show the magnetic field magnitudes; $B \approx \mathrm{constant}$ indicates the force-free current sheets. Middle panels show electron and ion $\beta$ profiles. Bottom panels show profiles of the field-aligned currents with an indication on the dominant ion species and estimations of the current sheet thickness in the ion inertial length. For the solar wind and Earth's lunar-distance magnetotail, we used ARTEMIS magnetic field \citep{Auster08:THEMIS} and plasma \citep{McFadden08:THEMIS} measurements \citep[see details of data processing procedure in][]{Artemyev19:jgr:solarwind}. For the Martian magnetotail, we used MAVEN magnetic field \citep{Connerney15:ssr} and plasma \citep{McFadden15:ssr,Halekas15:ssr} measurements \citep[see details of data processing procedure in][]{Artemyev17:jgr:mars}. For the Jovian magnetotail, we used Juno magnetic field \citep{Connerney17:ssr,Connerney17:science} and plasma \citep{McComas17:ssr,Kim20:juno:PS} measurements \citep[see details of data processing procedure in][]{Artemyev23:juno:arxiv}.  }
    \label{fig:space-observations}
\end{figure}

\blue{Although the plasma anisotropy may essentially affect current sheet configurations, taking into account all difficulties of the statistical determination of these effects, we perform a simplified categorization of current sheets using two plasma parameters: the Alfv\'en Mach number $M_A$ (i.e., the ratio of plasma flow speed to Alfv\'en speed) and the plasma beta $\beta$ (i.e., the ratio of plasma thermal pressure to magnetic field pressure).} Current sheet configurations from different space environments are located within different domains in this $(\beta, M_A)$ space. Tangential discontinuities ($B_n=0$) do not require contributions from the plasma dynamic pressure. These discontinuities exist in systems with either large $\beta$ (if the plasma pressure balances the magnetic field pressure) or small $\beta$ [if the current sheet configuration is magnetically force free without cross-field currents, ${\bf J}\times{\bf B}=0$; see Figure \ref{fig:scheme}(a)]. Such current sheets are often observed in the solar wind \citep[see discussion in][]{Neugebauer06,Artemyev19:grl:solarwind}, where they propagate with plasma flows and have $M_A \approx 0$ in the current sheet reference frame \citep[see examples of kinetic models of such current sheet configurations in][]{deKeyser96,Harrison09:prl,Allanson16:jpp,Neukirch20}. Rotational discontinuities are also observed in the solar wind ($\beta\sim1$) and are characterized by plasma flow gradients with $M_A\sim 1$ in the current sheet reference frame [\citet[e.g.,][]{deKeyser97,Haaland12,Paschmann13:angeo,Artemyev19:grl:solarwind}; see Figure \ref{fig:scheme}(b)]. Figure \ref{fig:scheme}(e) shows the parameter regime of such current sheets in $(\beta, M_A)$ space, whereas Figure \ref{fig:space-observations}(a) shows an example of a low-$\beta$, force-free current sheet in the solar wind. 

Current sheets in the shocked (magnetosheath) solar wind are characterized by lower $M_A$ [Figure \ref{fig:scheme}(e)] compared to that of the solar wind. The stress balance in these current sheets is significantly influenced by plasma pressure gradients, i.e., compressional current sheets with both parallel and transverse current density components \citep[e.g.,][]{Chaston&Travnicek20:magnetosheath_currents,Webster21}. Another $(\beta, M_A)$ domain, with similar $M_A$ and larger $\beta$, is found in Earth's distant magnetotail, where strong plasma flows may reach $M_A\geq 1$ and $\beta\in[10,100]$ \citep[e.g.,][]{Vasko15:jgr:cs,Artemyev17:jgr:mars}. These current sheets are mostly balanced by plasma pressure gradients. However, in the distant magnetotail and for $\beta \lesssim 10$ there have been previous observations of force-free current sheets with ${\bf J}\times{\bf B}=0$ \citep{Xu18:artemis_cs}. Figure \ref{fig:scheme}(e) shows the parameter domain of distant magnetotail current sheets, whereas Figure \ref{fig:space-observations}(b) shows an example of an almost force-free current sheet observed in the distant magnetotail. Both the magnetosheath and distant magnetotail current sheet configurations are characterized by a small but finite $B_n$, and thus the magnetic field line tension force $\propto B_n$ may be balanced by weak plasma flows or by plasma anisotropy. There is a series of models on such quasi-1D current sheets, with $B_n\ne 0$, that describe both large-$\beta$ \citep[see][]{Burkhart92TCS,Sitnov00,Sitnov06,Mingalev07,Zelenyi11PPR} and force-free \citep{Artemyev11:pop,Mingalev12,Vasko14:angeo_by} current sheets.

The distant magnetotail $(\beta, M_A)$ domain smoothly extends towards the lower $M_A$ with the decrease of the radial distance from the Earth. Near-Earth magnetotail current sheets are characterized by slightly higher $\beta>100$ and are quantified by the plasma pressure contribution to the stress balance \citep{Runov06,Artemyev11:jgr,Petrukovich15:ssr}. The important difference between solar wind/magnetosheath and magnetotail current sheets is their magnetic field line configuration; solar wind current sheets may be considered as 1D discontinuities, whereas the magnetotail current sheets are 2D [Figures \ref{fig:scheme}(a-d)]. Therefore, current sheet models with $B_n\ne0$ and low $M_A$ should include 2D plasma pressure gradients balancing the magnetic field line tension force \citep[see examples of such models in][]{SB02,Birn04,YL05,Sitnov&Merkin16}. 

In the $(\beta, M_A)$ space, the large-$\beta$ domain can be described by kinetic models of current sheets with $B_n=0$ (1D models) and $B_n\ne 0$ (2D models with plasma pressure gradients). Conversely, the low-$\beta$ domain with dominant field-aligned currents (force-free current sheets) can be described only by 1D kinetic models with $B_n=0$. Large $M_A$, low-$\beta$ (1D rotational discontinuities with $B_n\ne 0$) and low $M_A$, low-$\beta$ (2D force-free current sheet) domains have been analyzed only by fluid models \citep[e.g.,][]{Cowley78,Hilmer87,Tassi08,Lukin18}. The class of such 2D force-free current sheets is not limited to observations in the solar wind and distant Earth's magnetotail, but also includes current sheets observed in the cold plasma of the Martian magnetotail \citep[e.g.,][]{DiBraccio15,Artemyev17:jgr:mars} and in the low-density Jovian magnetotail \citep[e.g.,][]{Behannon81,Artemyev14:pss}. Figures \ref{fig:space-observations}(c,d) show examples of force-free current sheets observed in the Martian and Jovian magnetotails. There are not enough known integrals of motion to describe distribution functions of charged particles in such current sheet configurations \citep[see discussion in][]{Lukin22:arxiv}. Consequently, the absence of kinetic models of force-free current sheets with $B_n \ne 0$ significantly hinders the analysis of their stability, the process responsible for the magnetic reconnection onset and particle acceleration.

In this study, we develop a 2D kinetic force-free current sheet model. In Section \ref{sec:mhd-equilibrium}, we describe a magnetohydrodynamic (MHD) model of 2D force-free current sheets. In Section \ref{sec:setup}, we initialize self-consistent, particle-in-cell (PIC) simulations using currents and magnetic fields from the MHD equilibrium. We load particle distributions using drifting Maxwellians, which initially do not necessarily satisfy the time-stationary Vlasov equation. In Section \ref{sec:results}, we obtain kinetic equilibria by evolving these particle distribution functions using the self-consistent PIC simulations and demonstrate the existence of kinetic equilibria for 2D force-free current sheets. We describe the equilibrium distribution functions and the difference between ion- and electron-dominated 2D current sheets. In Section \ref{sec:conclusion}, we summarize the results and discuss their applications.

\section{MHD equilibria of force-free current sheets} \label{sec:mhd-equilibrium}
We consider the general case of 2D force-free equilibrium in which all quantities vary with coordinates $x$ and $z$ only (i.e., $\partial / \partial y = 0$), and, as is customary in this definition, the pressure gradient force contribution to the force balance is negligible, i.e., $\nabla P \sim 0$. The divergence-free magnetic field has the form \citep[e.g.,][]{low1988spontaneous}
\begin{equation} \label{eq:b-vec}
    \mathbf{B} = \left(-\frac{\partial A}{\partial z}, B_y, \frac{\partial A}{\partial x}\right) ,
\end{equation}
where $A \mathbf{e}_y$ is the vector potential in the $y$ direction, and $B_y$ is the magnetic field in the $y$ direction. The current density is
\begin{equation} \label{eq:j-vec}
    \mathbf{J} = \frac{c}{4 \pi} \nabla \times \mathbf{B} = \frac{c}{4 \pi} \left(-\frac{\partial B_y}{\partial z}, -\nabla^2 A, \frac{\partial B_y}{\partial x}\right) ,
\end{equation}
where $c$ is the speed of light, and $\nabla^2 = \partial^2/\partial x^2 + \partial^2/\partial z^2$ is the Laplacian. The force-free condition, $\mathbf{J} \times \mathbf{B} = 0$, is equivalent to $\mathbf{J} = \alpha \mathbf{B}$, where $\alpha = \alpha(x, z)$ is a scalar function. Comparing Equations \eqref{eq:b-vec} and \eqref{eq:j-vec}, the force-free condition requires that $B_y$ is a function of $A$ only, and that $B_y$ satisfies
\begin{equation} \label{eq:vec-pot}
    \nabla^2 A + B_y(A) \frac{\mathrm{d} B_y}{\mathrm{d} A} = 0 ,
\end{equation}
and the coefficient $\alpha$ is $(c / 4\pi) \mathrm{d}B_y / \mathrm{d}A$. Once $B_y(A)$ is given, we can solve Equation \eqref{eq:vec-pot} for $A(x, z)$ with appropriate boundary conditions.

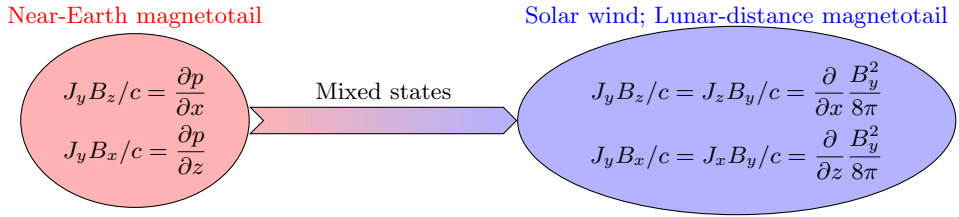
\begin{figure}[tphb]
    \centering
    \begin{tikzpicture}[
		signal arrow/.style={decorate,decoration={show path construction,  
				lineto code={
					\path let \p1 = ($(\tikzinputsegmentlast)-(\tikzinputsegmentfirst)$),
					\n1 = {int(mod(scalar(atan2(\y1,\x1))+360, 360))},
					\n2 = {veclen(\x1,\y1)}
					in
					(\tikzinputsegmentfirst) -- (\tikzinputsegmentlast)
					node[signal,midway,sloped,left color=red!30,right color=blue!30,draw,
					signal from=west, signal to=east,minimum width=\n2-\pgflinewidth,
					inner xsep=0pt,inner ysep=5pt,shading angle=\n1+90,
					anchor=center,#1]{};
		}}},
		signal arrow/.default= 
		]
		\path (0,0) node[process1] (A) {$\begin{aligned}
				J_y B_z / c = \frac{\partial p}{\partial x} \\
				J_y B_x / c = \frac{\partial p}{\partial z} 
			\end{aligned}$}
		(8,0) node[process2] (B) {$\begin{aligned}
				J_y B_z / c = J_z B_y / c = \frac{\partial}{\partial x} \frac{B_y^2}{8 \pi} \\
				J_y B_x / c = J_x B_y / c = \frac{\partial}{\partial z} \frac{B_y^2}{8 \pi}
			\end{aligned}$};
		\path[signal arrow] (A) node[anchor=south, yshift=32pt, color=red]{Near-Earth magnetotail} to node[anchor=south, yshift=5pt]{Mixed states}  (B) node[anchor=south, yshift=32pt, color=blue]{Solar wind; Lunar-distance magnetotail};
	\end{tikzpicture}
    \caption{Force balance of current sheets in different space environments. At one end of the spectrum, representing current sheets in the near-Earth magnetotail, magnetic tension force $J_y B_z /c$ and magnetic pressure force $J_y B_x / c \approx \partial (B_x^2 / 8 \pi) / \partial z$ are balanced by thermal pressure gradients $\partial p / \partial x$ and $\partial p / \partial z$, respectively. At the other end of the spectrum, representing current sheets in the solar wind and lunar-distance magnetotail, $J_y B_z /c$ and $J_y B_x / c$ are balanced by shears of $B_y^2$, i.e., $\partial (B_y^2 / 8 \pi) / \partial x$ and $\partial (B_y^2 / 8 \pi) / \partial z$, respectively. There may be a continuous spectrum of current sheets between these two ends \citep{yoon2023equilibrium}, in which $J_y B_z /c$ and $J_y B_x / c$ are balanced by a mixture of thermal pressure gradients and shears of $B_y^2$.}
    \label{fig:thermal-pressure-to-force-free}
\end{figure}

As shown in Figure \ref{fig:thermal-pressure-to-force-free}, for balancing force-free current sheets with low thermal and dynamic pressures (e.g., low $\beta$/low Mach number current sheets in the solar wind, magnetosheath, and lunar-distance magnetotail; see Figure \ref{fig:scheme}), the magnetic pressure $B_y^2$ plays an role analogous to the thermal pressure. Motivated by the functional form of pressure $p \propto \exp{\left[ 2 A / (\lambda B_0) \right]}$ describing thermal-pressure balanced current sheets \citep{LP82}, we adopt
\begin{equation}
	B_y(A) = B_0 \exp\left(\frac{A}{\lambda B_0}\right),
\end{equation}
for force-free current sheets, so that Equation \eqref{eq:vec-pot} describing magnetic field lines in the $x$-$z$ plane (i.e., contours of $A$) resembles that of the Lembege-Pellat current sheet \citep[e.g.,][]{Tassi08,Lukin18}:
\begin{equation}\label{eq:Lembege-Pellat-general}
	\nabla^2 A = -\frac{B_0}{\lambda} \exp\left(\frac{2 A}{\lambda B_0}\right) ,
\end{equation}
where $\lambda$ is the current sheet thickness, and $B_0$ is the magnetic field at large $\vert z \vert$.

Because the scale length along current sheets is much larger than that across current sheets, we assume $A = A(\varepsilon x, z)$ to be weakly nonuniform in the $x$ direction ($\varepsilon$ being a small parameter). Equation \eqref{eq:Lembege-Pellat-general} can be approximated as
\begin{equation}\label{eq:Lembege-Pellat-magnetotail}
	\frac{\partial^2 A}{\partial z^2} = -\frac{B_0}{\lambda} \exp\left(\frac{2 A}{\lambda B_0}\right) ,
\end{equation}
which is accurate up to order $\varepsilon$. The boundary condition is
\begin{equation}\label{eq:bc-magnetotail}
	\frac{\partial A}{\partial z}\bigg\vert_{z = 0} = 0, \hspace{15pt} A\big\vert_{z = 0} = \varepsilon B_0 x ,
\end{equation}
which is equivalent to $B_x (z = 0) = 0$ and $B_z (z = 0) = \varepsilon B_0$. The solution of $A$ is
\begin{equation}
	A = - \lambda B_0 \ln\left[\cosh\left(\frac{z}{\lambda F(x)}\right) \cdot F(x)\right] ,
\end{equation}
where $F(x) = \exp\left(-\varepsilon x / \lambda\right)$. Thus the three components of the magnetic field are
\begin{equation} \label{eq:b-vec-LP}
	\begin{split}
		B_x &= B_0 \tanh\left(\frac{z}{\lambda F(x)}\right) F^{-1}(x) , \\
		B_y &= B_0 \left[\cosh\left(\frac{z}{\lambda F(x)}\right) \cdot F(x)\right]^{-1} , \\
		B_z &= \varepsilon B_0 \left[1 - \frac{z}{\lambda F(x)} \tanh\left(\frac{z}{\lambda F(x)}\right)\right] ,
	\end{split}
\end{equation}
where $B_z \neq 0$ describes rotational discontinuities. The current density is
\begin{equation} \label{eq:j-vec-LP}
	\mathbf{J} = \alpha \mathbf{B} = \frac{c \mathbf{B}}{4 \pi \lambda} \left[\cosh\left(\frac{z}{\lambda F(x)}\right) \cdot F(x)\right]^{-1} .
\end{equation}
The direction of magnetic field rotates from pointing in $+x$ at the $z>0$ half-space ($z / \lambda \gg 1$) to pointing in $+y$ around the neutral (equatorial) plane ($-1 \lesssim z / \lambda \lesssim 1$), and further to pointing in $-x$ at the $z<0$ half-space   ($z / \lambda \ll -1$). The magnitude of magnetic field, $(B_x^2 + B_y^2 + B_z^2)^{1/2} = B_0 [F^{-1}(x) + \mathcal{O}(\varepsilon)]$, has a weak dependence on $x$ and is roughly a constant at a given $x$ location. The magnitude of current density, $(J_x^2 + J_y^2 + J_z^2)^{1/2} \propto \cosh^{-1}[z / (\lambda F(x))]$, is concentrated in a layer $\vert z \vert < \lambda F(x)$. \blue{The current density profile [Equation \eqref{eq:j-vec-LP}], together with the constant plasma pressure profile, constrains the initial particle distribution functions.}

\section{Computational setup} \label{sec:setup}
Searching for kinetic equilibria of 2D force-free current sheets, we initialize our simulated current sheets from the results of an MHD equilibrium. Importantly, this may not satisfy the time-stationary Vlasov equation because the initial particle distributions are simply drifting Maxwellians. We simulate their relaxation toward a certain kinetic equilibrium (if any) based on a massively parallel PIC code \citep{Pritchett01,Pritchett05:driven}. Our simulations have two dimensions $(x, z)$ in configuration space and three dimensions $(v_x, v_y, v_z)$ in velocity space. The results are presented in normalized units: magnetic fields are normalized to $B_0$, lengths to the ion inertial length $d_i = c / (4 \pi n_0 e^2 / m_i)^{1/2}$, time to the reciprocal of the ion gyrofrequency $\omega_{ci}^{-1} = m_i c / (e B_0)$, velocities to the Alfv\'en velocity $v_\mathrm{A} = B_0 / (4 \pi n_0 m_i)^{1/2}$, electric fields to $v_\mathrm{A} B_0 / c$, and energies to $m_i v_\mathrm{A}^2$. Here $n_0$ is the plasma density, $m_i$ is the ion mass, and $e$ is the elementary charge. The computational domain is $[-64 \leq x / d_i \leq 0] \times [-8 \leq z / d_i \leq 8]$ with a cell length $\Delta x = d_i / 32$. The time step is $\Delta t = 0.001\, \omega_{ci}^{-1}$. The ion-to-electron mass ratio is $m_i / m_e = 100$. The normalized speed of light is $c / v_\mathrm{A} = 20$, which gives the ratio of electron plasma frequency to electron gyrofrequency $\omega_{pe} / \omega_{ce} = 2$. The reference density $n_0$ is represented by $1929$ particles. The total number of particles is $1.6 \times 10^9$ including ions and electrons in each simulation.

\begin{figure}[tphb]
    \centering
    \includegraphics[width=0.6\textwidth]{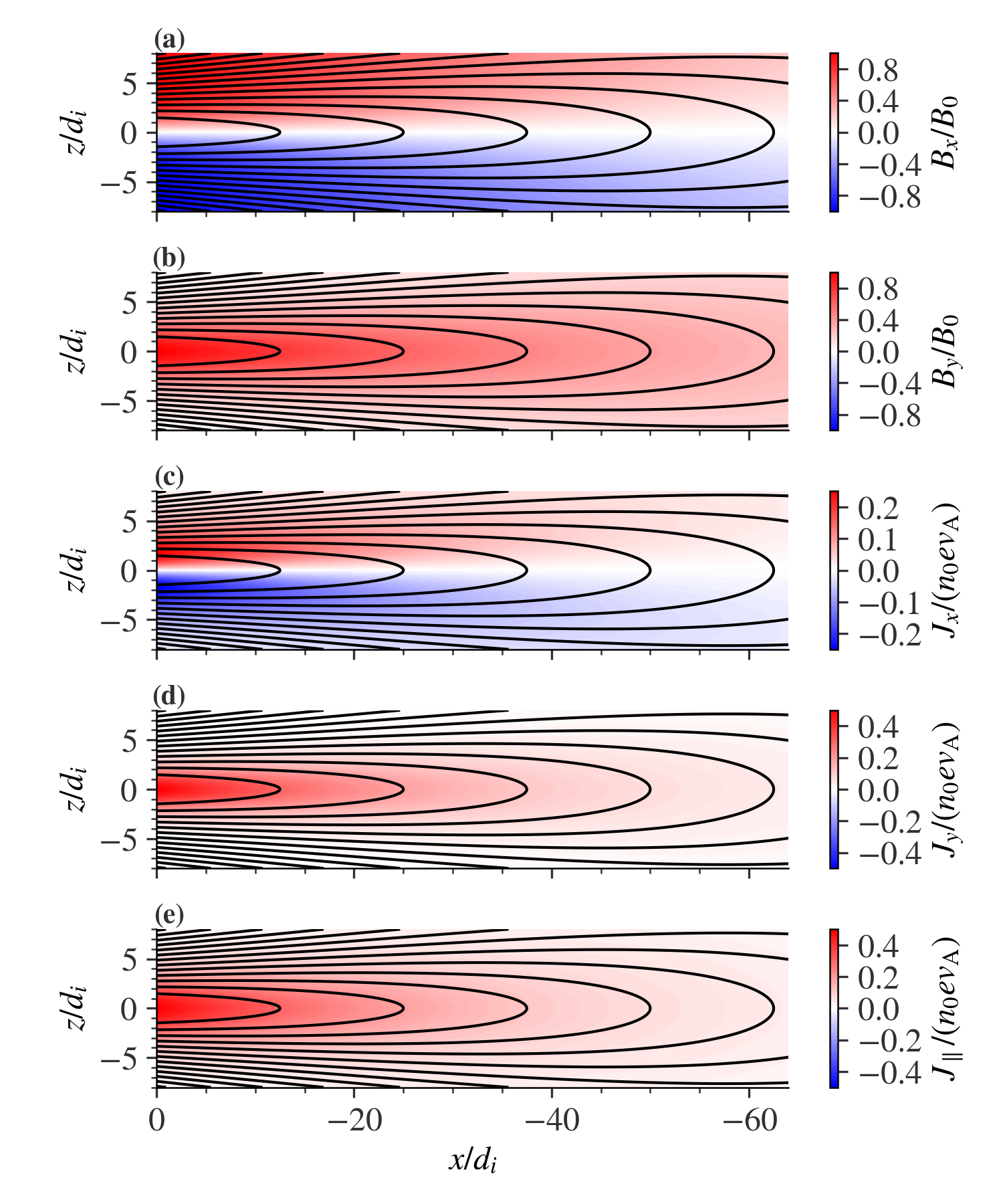}
    \caption{MHD equilibrium of a force-free current sheet. (a) Magnetic field in the $x$ direction $B_x$. (b) Magnetic field in the $y$ direction $B_y$. (c) Current density in the $x$ direction $J_x$. (d) Current density in the $y$ direction $J_y$. (e) Current density in the parallel direction $J_\parallel$.}
    \label{fig:init-config}
\end{figure}

Figure \ref{fig:init-config} shows the initial magnetic field configuration and current density, which are determined by Equations \eqref{eq:b-vec-LP} and \eqref{eq:j-vec-LP} using the input parameters $\varepsilon = 0.04$ and $\lambda = 2 d_i$. There are a few degrees of freedom in choosing the initial particle distribution functions: (1) While the current density is known a priori, the proportion of ion and electron currents is left unspecified; (2) Electron and ion temperatures and plasma $\beta$ vary across different space environments (e.g., Figure \ref{fig:space-observations}); (3) Specific forms of the velocity distribution functions are left undetermined. To this end, we choose initial electron and ion distribution functions as drifting Maxwellians
\begin{equation}
    f_{s} (\mathbf{v}, x, z) = \frac{n_0}{(2 \pi T_{s} / m_s)^{3/2}} \exp\left[-\frac{m_s \left(\mathbf{v} - \mathbf{v}_{d s}(x, z)\right)^2}{2 T_s}\right] ,
\end{equation}
where the subscript $s = i, e$ represents ions and electrons, $\mathbf{v}_{d s}(x, z)$ is the position-dependent drift velocity, and $T_s$ is the temperature. Note that $T_i$, $T_e$, and $n_0$ are constants throughout the domain so that the initial plasma pressure $n_0 (T_i + T_e)$ is a constant, as required by the force-free condition. The simulations allow either ions or electrons to be the main current carrier, while keeping the relative drift between them commensurate with the current density [Figures \ref{fig:init-config}(c), \ref{fig:init-config}(d), and \ref{fig:init-config}(e)]. Additionally, we vary particle temperatures to search for kinetic equilibrium in different plasma beta regimes, where the plasma beta is $\beta = 2(T_i + T_e) /m_i v_{\mathrm{A}}^2$. The detailed parameters for particle distribution functions in the six runs are listed in Table \ref{tab:init-dist-params}. These parameters cover the $\beta$ and $T_i/T_e$ ranges of force-free current sheets observed in the solar wind and planetary magnetotails (see Figure \ref{fig:space-observations}).

\begin{table}[]
    \centering
    \begin{tabular}{c|c|c|c|c|c}
    \tableline
        Run NO. & $\mathbf{v}_{di}$      & $\mathbf{v}_{de}$       & $\beta_i = 2 T_i / m_i v_\mathrm{A}^2$      & $\beta_e = 2 T_e / m_i v_\mathrm{A}^2$      & $\beta = \beta_i + \beta_e$ \\
        \tableline
        1A      & $\mathbf{J} / (n_0 e)$ & $0$                     & $0.05$         & $0.05$         & $0.1$ \\
        1B      & $0$                    & $-\mathbf{J} / (n_0 e)$ & $0.05$         & $0.05$         & $0.1$\\
        \tableline
        2A      & $\mathbf{J} / (n_0 e)$ & $0$                     & $5/6$          & $1/6$          & $1$ \\
        2B      & $0$                    & $-\mathbf{J} / (n_0 e)$ & $5/6$          & $1/6$          & $1$ \\
        \tableline
        3A      & $\mathbf{J} / (n_0 e)$ & $0$                     & $25/3$         & $5/3$          & $10$ \\
        3B      & $0$                    & $-\mathbf{J} / (n_0 e)$ & $25/3$         & $5/3$          & $10$ \\
        \tableline
    \end{tabular}
    \caption{Parameters for particle distribution functions in the six PIC runs. The first digit in Run NO., `1', `2', `3', denotes three plasma betas, $\beta = 0.1, 1, 10$, respectively. The second digit in Run NO., `A', `B', denotes current carriers as ions and electrons, respectively. The current density $\mathbf{J}$ comes from Equation \eqref{eq:j-vec-LP}, and is visualized in Figures \ref{fig:init-config}(c) and \ref{fig:init-config}(d).}
    \label{tab:init-dist-params}
\end{table}

In our simulations, the electromagnetic fields are advanced in time by integrating Maxwell's equations using a leapfrog scheme. These fields are stored on the Yee grid. The relativistic equations of motion for ions and electrons are integrated in time using a leapfrog scheme with a standard Boris push for velocity update. The conservation of charge is ensured by applying a Poisson correction to the electric fields \citep{marder1987method,langdon1992enforcing}.

For particles crossing the $x$ boundaries, we take advantage of the symmetry between $z>0$ and $z<0$ [Figure \ref{fig:init-config}], so that particles exiting the system at a location $z$ with velocity $(v_x, v_y, v_z)$ are reinjected into the system at the conjugate location $-z$ with velocity $(-v_x, v_y, v_z)$. This is equivalent to an open boundary condition for particles because the injected particle distribution matches that at one cell interior to the boundary at all simulation times. At the $z$ boundaries, particles striking the boundary are reflected into the system with $v_z = - v_z$.

For fields at the $x$ boundaries, the guard values of the tangential magnetic fields are determined by
\begin{equation}
    \delta B^n_{y, g} = \delta B^n_{y, i1} , \hspace{15pt} \delta B^n_{z, g} = \delta B^{n-1}_{z, i1} (2 - \Delta x / c \Delta t) + \delta B^{n-2}_{z, i2} (\Delta x / c \Delta t - 1) ,
\end{equation}
where the superscript indicates the time level, and the subscripts $g, i1, i2$ indicate the guard point, first interior point and second interior point, respectively. This boundary condition for $\delta B_z$ ensures that the magnetic flux can freely cross the $x$ boundaries \citep{pritchett2005newton}. The guard values of the normal electric field $\delta E_x$ at the $x$ boundaries are determined by $\delta E_{x, g}^n = \delta E_{x, i1}^n$. Similarly, at the $z$ boundaries, the two components of the tangential magnetic fields in the guard point are determined by $\delta B_{x, g}^n = \delta B_{x, i1}^n$ and $\delta B_{y, g}^n = \delta B_{y, i1}^n$; The normal electric field $\delta E_z$ in the guard point are determined by $\delta E_{z, g}^n = \delta E_{z, i1}^n$. Unlike the driven simulations that adds magnetic flux to the simulation box, no external $E_y$ is applied at the $z$ boundaries. 

\section{Results} \label{sec:results}
We track the evolution of the initially force-free current sheets in the six runs up to $t = 180\, \omega_{ci}^{-1}$, at which time the macroscopic states (e.g., electric and magnetic fields, currents, pressures) of the system are quasi-stationary. Below we examine if the configurations satisfy $\mathbf{J} \times \mathbf{B} = 0$ at the end of the simulations. We further investigate how the particle velocity distributions deviate from the initial Maxwellians and whether or not the system reaches a kinetic equilibrium.

\subsection{Initially ion-dominated current sheets}
\begin{figure}[tphb]
    \centering
    \includegraphics[width=0.6\textwidth]{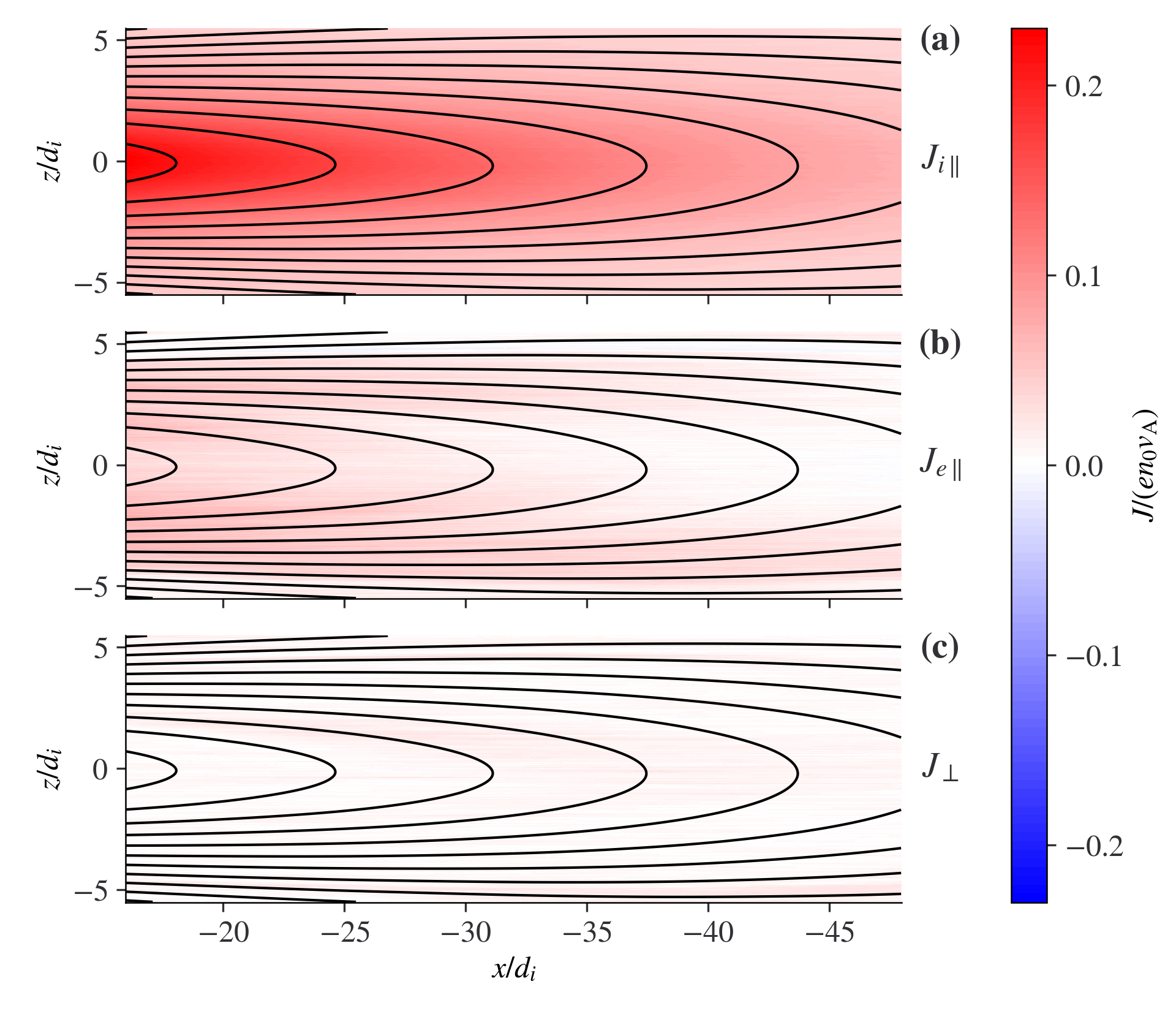}
    \caption{Current density for the ion-dominated force-free current sheet with plasma beta $\beta = 1$ in Run 2A. The snapshot is taken at $t = 180\, \omega_{ci}^{-1}$ in the simulation. (a) Ion field-aligned currents. (b) Electron field-aligned currents. (c) Total perpendicular currents.}
    \label{fig:current-case1}
\end{figure}

When ions initially carry entirely the (field-aligned) currents (Runs 1A, 2A, and 3A), electrons are accelerated by transient parallel electric fields to form field-aligned currents [e.g., Figure \ref{fig:current-case1}(b)], while ions are slightly decelerated by such electric fields [e.g., comparing Figures \ref{fig:current-case1}(a) and \ref{fig:init-config}(e)]. These electron currents can be a significant fraction (e.g., $\gtrsim 1/3$) of ion currents in the off-equatorial region ($1 < \vert z \vert < 5$). Almost all currents remain field aligned [see $J_\perp \sim 0$ in Figure \ref{fig:current-case1}(c)], implying that a force-free configuration may indeed be obtained.

\begin{figure}[tphb]
    \centering
    \includegraphics[width=0.6\textwidth]{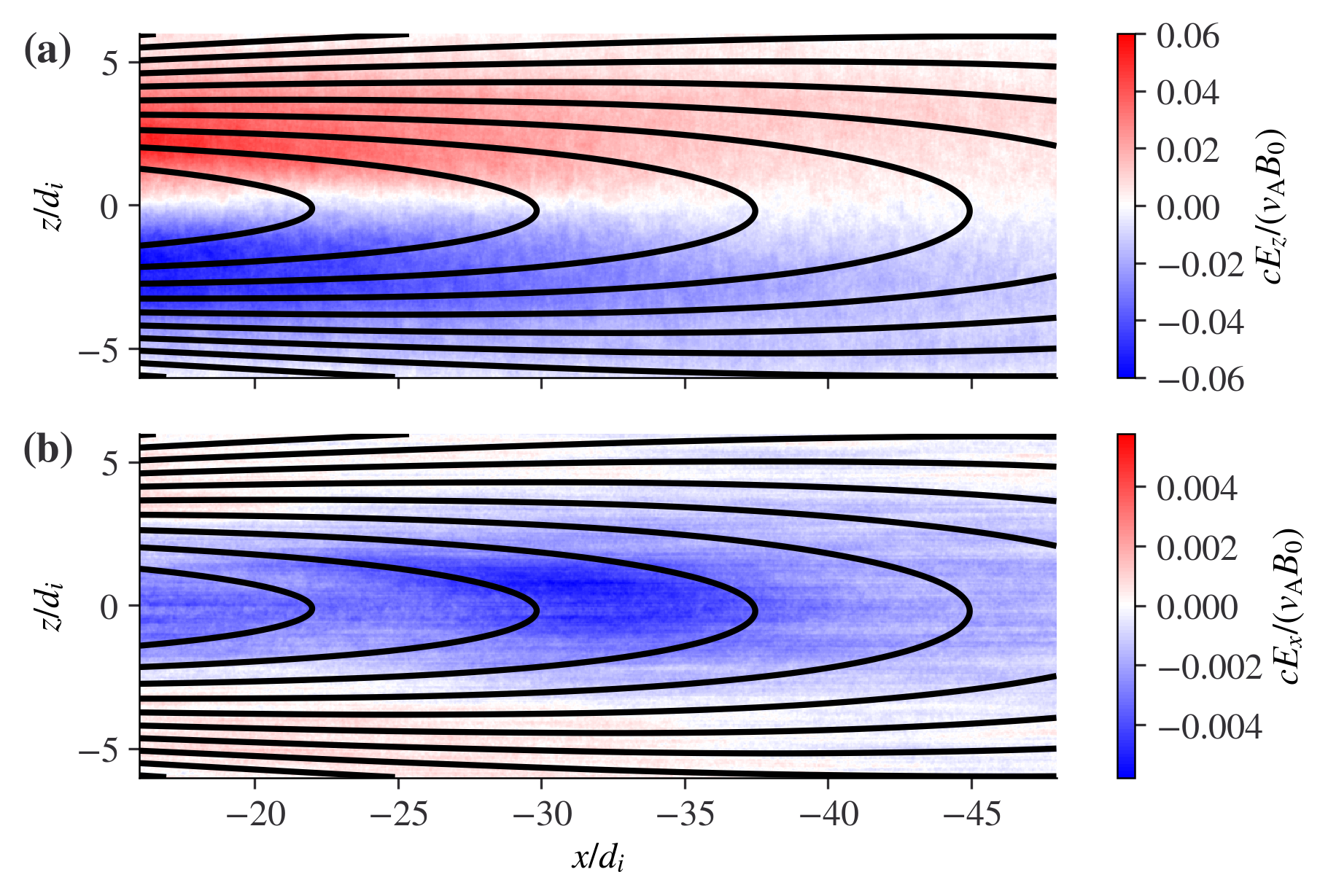}
    \caption{Electrostatic field for the ion-dominated force-free current sheet with plasma beta $\beta = 1$ in Run 2A. The snapshot is taken at $t = 180\, \omega_{ci}^{-1}$ in the simulation. (a) Electric field in the $z$ direction $E_z$. (b) Electric field in the $x$ direction $E_x$. The magnitude of $E_x$ is much smaller than that of $E_z$.}
    \label{fig:efield}
\end{figure}

During the relaxation of ion-dominated current sheets, electrostatic fields are generated and remain present in the late equilibrium states. These electrostatic fields are perpendicular to the magnetic field: At $\vert z \vert > 0$, they points away from the equatorial plane [Figure \ref{fig:efield}(a)]; Near the equator, they points in the $-x$ direction (i.e., tailward) [Figure \ref{fig:efield}(b)]. Here the $E_x$ component is much weaker than the $E_z$ component. The electrostatic fields arise due to the decoupling of the unmagnetized ions and magnetized electrons \citep{Schindler12}, which can be derived from the ordering among ion thermal gyroradius, current sheet thickness, and electron thermal gyroradius
\begin{equation}
    \rho_i : \lambda : \rho_e  = d_i \sqrt{\frac{\beta_i}{2}} : 2 d_i : d_i \sqrt{\frac{\beta_e}{2} \frac{m_e}{m_i}} = \sqrt{\frac{\beta_i}{8}} : 1 : \sqrt{\frac{\beta_e}{8} \frac{m_e}{m_i}} .
\end{equation}
Taking Run 2A for example, we have $\rho_i : \lambda : \rho_e = 0.32 : 1 : 0.01$. As the plasma beta is lowered, the electrostatic field decreases -- due to stronger magnetization of ions, which is shown in Figures \ref{fig:fbx} and \ref{fig:fbz} below.

To show the detailed force balance of these current sheet configurations, we decompose the forces in parallel and perpendicular directions with respect to local magnetic fields. The force balance in the parallel direction is rapidly established because particles move freely along field lines. The unit vectors in the two perpendicular directions are defined as $\mathbf{e}_{\perp 1} = \hat{\mathbf{z}} \times \hat{\mathbf{b}}$ and $\mathbf{e}_{\perp 2} = \hat{\mathbf{b}} \times (\hat{\mathbf{z}} \times \hat{\mathbf{b}})$, where $\hat{\mathbf{b}}$ and $\hat{\mathbf{z}}$ are the unit vectors along the magnetic field and $z$ direction, respectively. It is worth noting that $\mathbf{e}_{\perp 1}$ is roughly along the $\pm y$ direction far above/below the equator, and along the $-x$ direction at the equator, while $\mathbf{e}_{\perp 2}$ is roughly along the $+z$ direction all over the domain.

At the equator, the electrostatic fields, $\mathbf{E}_{\perp 1}$ (i.e., along the $-x$ direction), cause a drift $c \mathbf{E}_{\perp 1} \times \mathbf{b} / \vert \mathbf{B} \vert$ of both ions and electrons [Figures \ref{fig:fbx}(b-c), \ref{fig:fbx}(h-i), and \ref{fig:fbx}(n-o)], which does not produce net currents [Figures \ref{fig:fbx}(a), \ref{fig:fbx}(g), and \ref{fig:fbx}(m)]. In addition, the pressure gradient $(\nabla \cdot \mathbf{P})_{\perp 1}$ is about zero in the two cases, with $\beta = 0.1$ [Run 1A; see Figure \ref{fig:fbx}(a)] and $\beta = 1$ [Run 2A; see Figure \ref{fig:fbx}(g)]. Thus, we obtain $(\mathbf{J} \times \mathbf{B})_{\perp 1} / c = (\nabla \cdot \mathbf{P})_{\perp 1} = 0$ in the $\mathbf{e}_{\perp 1}$ direction. In the case with $\beta = 10$, however, the current sheet gets compressed and rarefied periodically in the $x$ direction. Correspondingly, $(\nabla \cdot \mathbf{P})_{\perp 1}$ oscillates around zero and acts as a restoring force [Run 3A; see Figure \ref{fig:fbx}(m)]. Such oscillations are still localized around the equilibrium $(\mathbf{J} \times \mathbf{B})_{\perp 1} / c = (\nabla \cdot \mathbf{P})_{\perp 1} = 0$ when performing a spatial average over $x$.

Along the $\mathbf{e}_{\perp 2}$ direction (i.e., roughly the $z$ direction), both the pressure gradient $(\nabla \cdot \mathbf{P})_{\perp 2}$ and Lorentz force $(\mathbf{J} \times \mathbf{B})_{\perp 2} / c$ vanish [Figures \ref{fig:fbz}(a), \ref{fig:fbz}(g), \ref{fig:fbz}(m)]. The dominant component of the electrostatic fields, $\mathbf{E}_{\perp 2}$ (along the $\pm z$ directions), leads to a drift $c \mathbf{E}_{\perp 2} \times \mathbf{b} / \vert \mathbf{B} \vert$ (roughly in the $+y$ direction) of both ions and electrons [Figures \ref{fig:fbz}(b-c), \ref{fig:fbz}(h-i), and \ref{fig:fbz}(n-o)], which does not give net currents. All aforementioned perpendicular drift velocities are small, when compared to ion and electron parallel flow velocities that carry currents [e.g., Figure \ref{fig:current-case1}].

\blue{In all cases of ion-dominated current sheets, the $\mathbf{E} \times \mathbf{B}$ drift of ions and electrons does not give net perpendicular currents. The total electric currents are field-aligned. In addition, the plasma pressure gradients are approximately zero. Thus, the force-free configurations of the ion-dominated current sheets are obtained.}

\begin{figure}[tphb]
    \centering
    \includegraphics[width=0.9\textwidth]{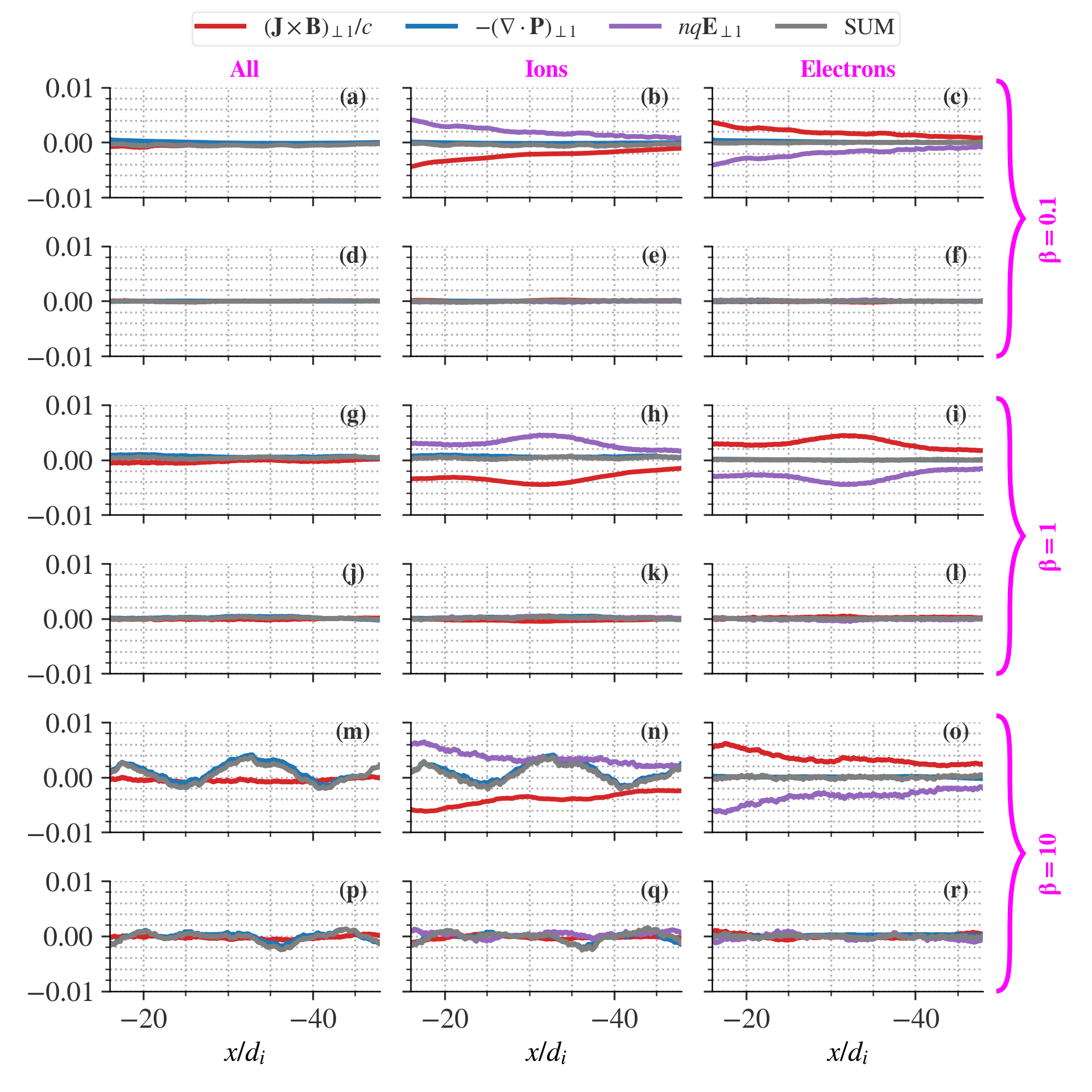}
    \caption{Force balance in the $\mathbf{e}_{\perp 1} = \hat{\mathbf{z}} \times \hat{\mathbf{b}}$ direction at the equator for different plasma betas. Note $\mathbf{e}_{\perp 2}$ is along the $-x$ direction at the equator. All the snapshots are taken at $t = 180\, \omega_{ci}^{-1}$ in the simulations. The volume forces are averaged in the range $-1 \leq z / d_i \leq 1$. The rows from top to bottom represent Runs 1A, 1B, 2A, 2B, 3A, and 3B, respectively. The columns from left to right represent the force balance from the point view of single fluid, ion fluid, and electron fluid, respectively.}
    \label{fig:fbx}
\end{figure}

\begin{figure}[tphb]
    \centering
    \includegraphics[width=0.9\textwidth]{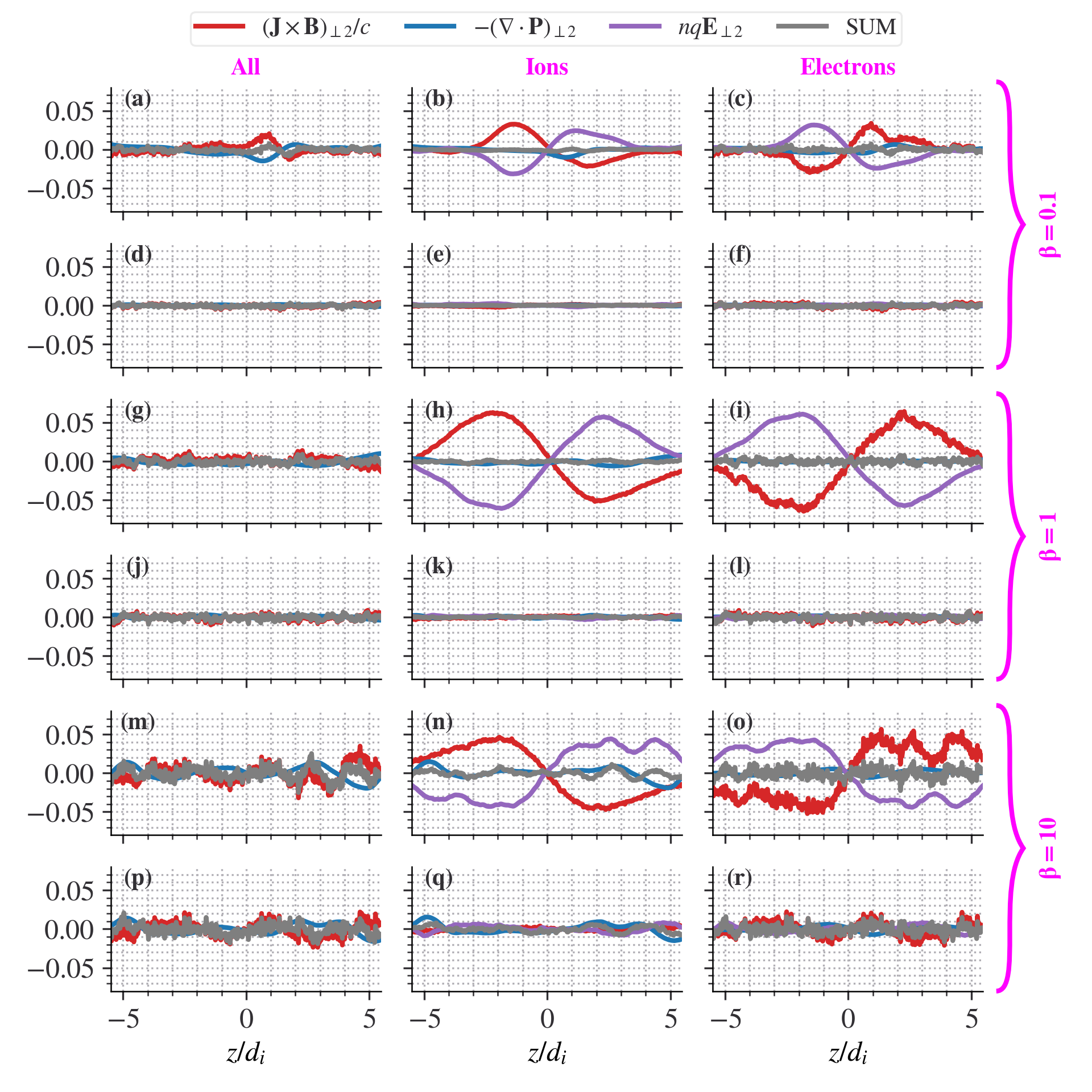}
    \caption{Force balance in the $\mathbf{e}_{\perp 2} = \hat{\mathbf{b}} \times (\hat{\mathbf{z}} \times \hat{\mathbf{b}})$ direction for different plasma betas. Note $\mathbf{e}_{\perp 2}$ is approximately along the $+z$ direction. All the snapshots are taken at $t = 180\, \omega_{ci}^{-1}$ in the simulations. The volume forces are averaged in the range $-16 \leq x / d_i \leq -12$. The format is the same as Figure \ref{fig:fbx}.}
    \label{fig:fbz}
\end{figure}

\subsection{Initially electron-dominated current sheets}
\begin{figure}[tphb]
    \centering
    \includegraphics[width=0.6\textwidth]{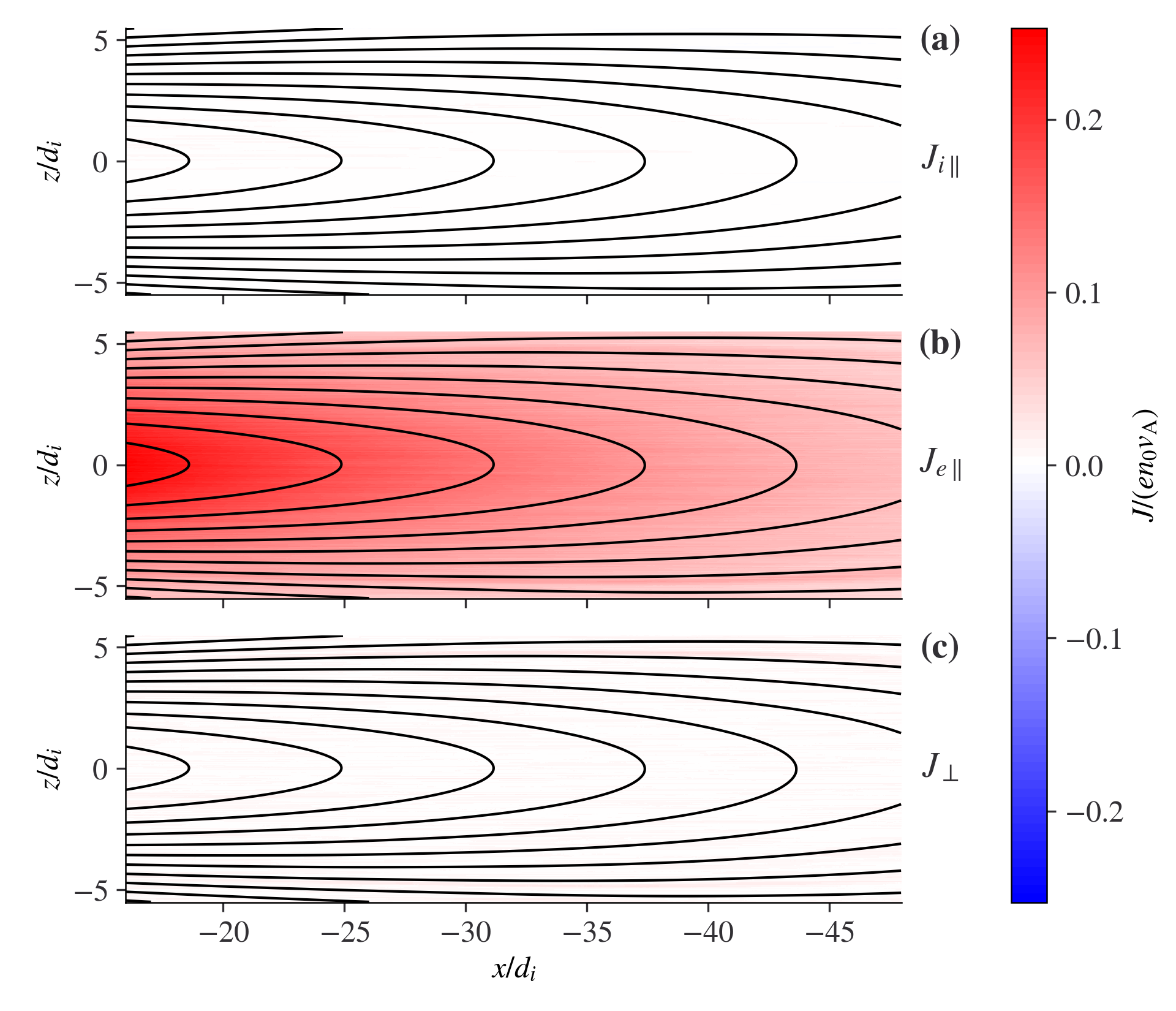}
    \caption{Current density for the electron-dominated force-free current sheet with plasma beta $\beta = 1$ in Run 2B. The snapshot is taken at $t = 180\, \omega_{ci}^{-1}$ in the simulation. (a) Ion field-aligned currents. (b) Electron field-aligned currents. (c) Total perpendicular currents.}
    \label{fig:current-case2}
\end{figure}

When electrons initially carry entirely the (field-aligned) currents (Runs 1B, 2B, and 3B), there is no change in the macroscopic states of the system at the end of simulations [e.g., Figure \ref{fig:current-case2}]: The currents remain field aligned, and are carried by electrons only. Compared to the ion-dominated current sheets, the relaxation of electron-dominated current sheets does not alter the macroscopic states: neither the proportion of electron and ion currents is changed, nor do the electrostatic fields develop. In all three cases with different plasma betas, the system remains in a force-free configuration, with no plasma pressure gradients [Figures \ref{fig:fbx}(d-f), \ref{fig:fbx}(j-l), \ref{fig:fbx}(p-r), \ref{fig:fbz}(d-f), \ref{fig:fbz}(j-l), \ref{fig:fbz}(p-r)]. Despite the system being identical to the initial ones from the fluid point of view, the underlying velocity distribution functions evolve substantially from the initial Maxwellians to reach the Vlasov equilibrium, shown below.

\subsection{Reaching kinetic equilibrium or not?}
Since the force-free equilibria are reestablished toward the end of simulations for both ion- and electron-dominated current sheets from the fluid point of view, a logical next step would be to further examine if such current sheets reach the kinetic equilibrium. Because all the currents are field aligned, we integrate the full distribution function $f_s(x, z, v_\parallel, v_\perp, \phi, t)$ of species $s$ to obtain the reduced distribution $g_s(x, v_\parallel, t)$ at the equator as a function of $x$, $v_\parallel$, and $t$:
\begin{equation}
    g_s(x, v_\parallel, t) = \int_{-d_i}^{d_i} \mathrm{d} z \int_{0}^{\infty} \mathrm{d} v_\perp \int_{0}^{2 \pi} \mathrm{d} \phi\, f_s(x, z, v_\parallel, v_\perp, \phi, t) ,
\end{equation}
where $v_\parallel$ and $v_\perp$ are the parallel and perpendicular velocities, respectively, and $\phi$ is the gyrophase.

\begin{figure}[tphb]
    \centering
    \includegraphics[width=0.9\textwidth]{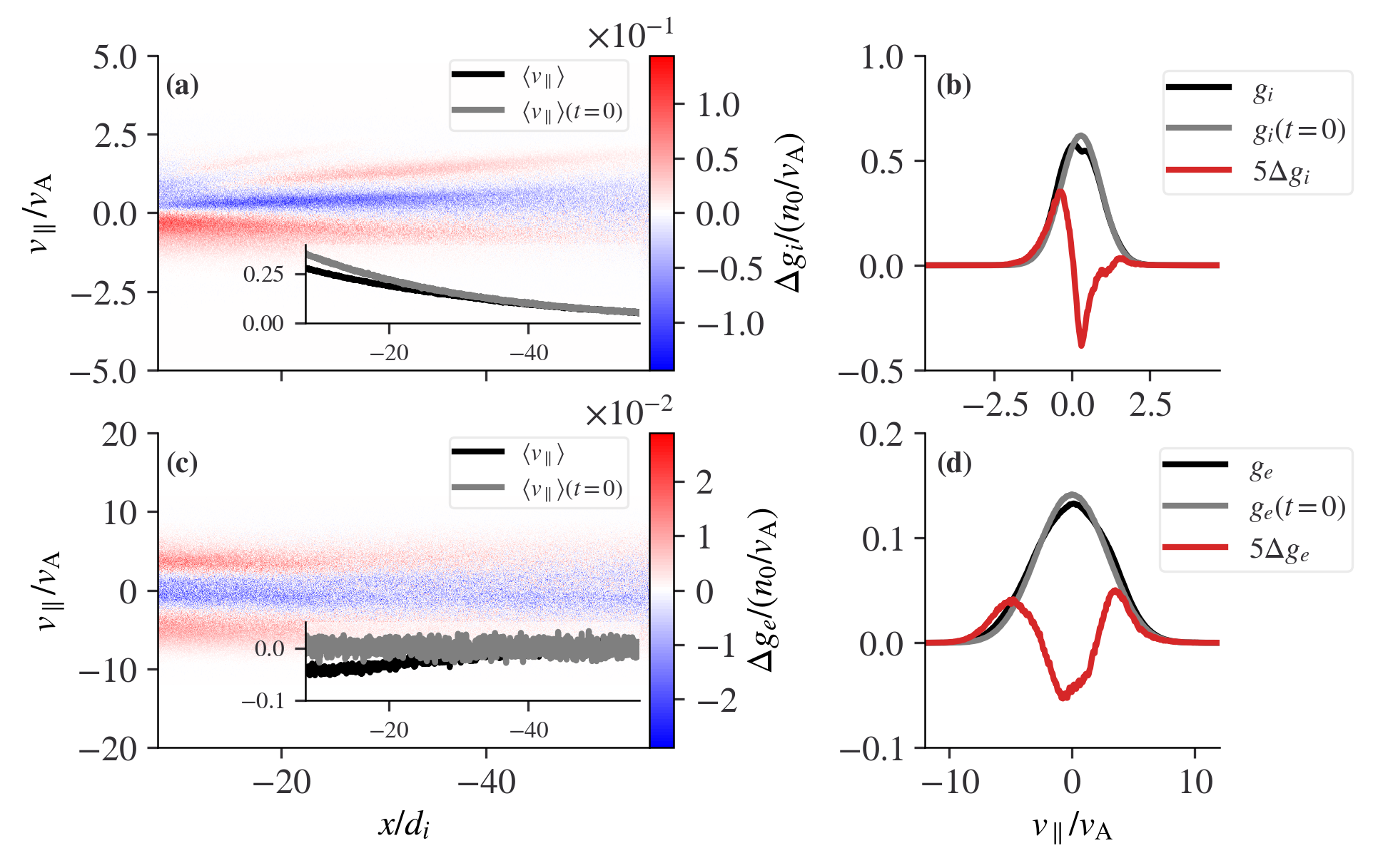}
    \caption{Phase space distributions for the ion-dominated force-free current sheet with plasma beta $\beta = 1$ in Run 2A. (a) The difference between the reduced ion distribution and the initial Maxwellian $\Delta g_i$. The inset plot shows the mean ion flow velocity as a function of $x$ at $t=0$ (gray line) and $t = 180\, \omega_{ci}^{-1}$ (black line). (b) A cut of the final ion distribution $g_i (x, v_\parallel, t = 180\, \omega_{ci}^{-1})$ (black line), $g_i (x, v_\parallel, t = 0)$ (gray line), and their difference $\Delta g_i$ multiplied by $5$, for better visibility (red line) between $-16 \leq x / d_i \leq -12$. (c) The different between the reduced electron distribution and the initial Maxwellian $\Delta g_e$. The inset plot shows the mean electron flow velocity as a function of $x$ at $t=0$ (gray line) and $t = 180\, \omega_{ci}^{-1}$ (black line). (d) A cut of the final electron distribution $g_e (x, v_\parallel, t = 180\, \omega_{ci}^{-1})$ (black line), $g_e (x, v_\parallel, t = 0)$ (gray line), and their difference $\Delta g_e$ multiplied by $5$, for better visibility (red line) between $-16 \leq x / d_i \leq -12$.}
    \label{fig:vphase_space_case1}
\end{figure}

Figure \ref{fig:vphase_space_case1} shows the difference between the reduced distributions at the end of simulations and the initial Maxwellians, $\Delta g_s = g_s(x, v_\parallel, t= 180\, \omega_{ci}^{-1}) - g_s(x, v_\parallel, t= 0)$, from Run 2A as an example. The ion phase space density is transported toward velocities both above and below the mean ion flow velocity [Figure \ref{fig:vphase_space_case1}(a)]. Consequently, the ion velocity distribution becomes wider and less peaked, compared to the initial Maxwellian [Figure \ref{fig:vphase_space_case1}(b)]. The mean ion flow velocity is slowed more on the earthward side than the tailward side [see the inset of Figure \ref{fig:vphase_space_case1}(a)]. Correspondingly, electrons are accelerated on average in the $v_\parallel < 0$ direction to develop an electron current to compensate the reduction of the ion current [see the inset of Figure \ref{fig:vphase_space_case1}(c)]. Such changes of mean flow velocities are not apparent in the phase space density plot because the thermal velocities are much larger than the mean flow velocities. The peak of the electron phase space density is transported toward larger velocities in both $v_\parallel > 0$ and $v_\parallel < 0$ [Figures \ref{fig:vphase_space_case1}(c) and \ref{fig:vphase_space_case1}(d)].

\begin{figure}[tphb]
    \centering
    \includegraphics[width=0.9\textwidth]{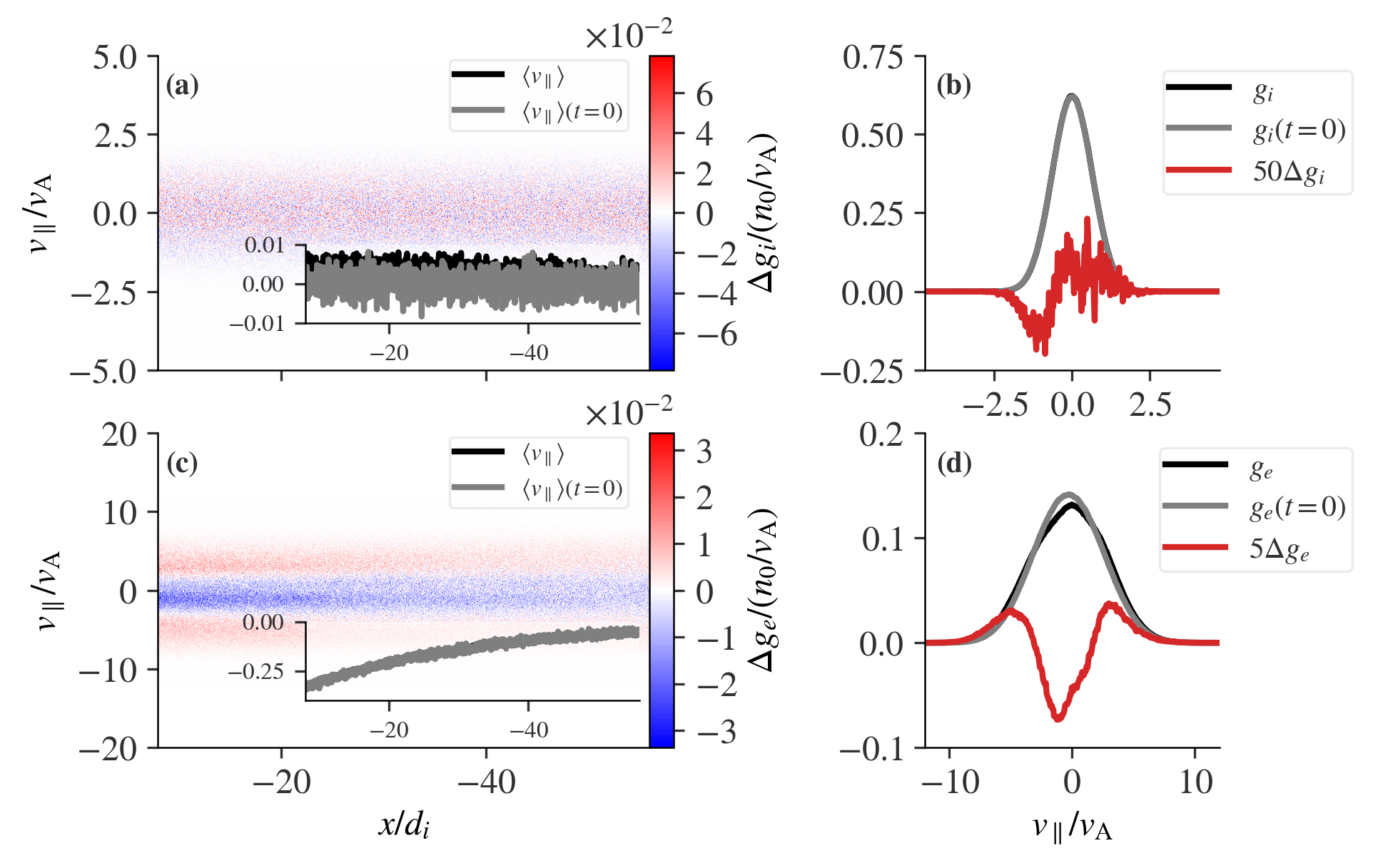}
    \caption{Phase space distributions for the electron-dominated force-free current sheet with plasma beta $\beta = 1$ in Run 2B. The format is the same as Figure \ref{fig:vphase_space_case1}. \blue{Although the profiles of particle drift velocities are almost the same as their initial [see the insets of (a) and (c)], the particle distribution functions evolve substantially from the initial drift Maxwellians [see panels (b) and (d)].}}
    \label{fig:vphase_space_case2}
\end{figure}

The macroscopic states of electron-dominated current sheets do not differ from their initial configurations, as seen in Figure \ref{fig:current-case2}. The electron reduced distribution function $g_e(x, v_\parallel)$, however, shows systematic deviations from the initial Maxwellians, as displayed in Figure \ref{fig:vphase_space_case2}. Such deviations are similar to those in ion-dominated current sheets. The peak of the electron velocity distributions is reduced and redistributed toward larger velocities of $v_\parallel > 0$ and $v_\parallel < 0$ (but without changes in mean electron flow velocities) [Figures \ref{fig:vphase_space_case2}(c) and \ref{fig:vphase_space_case2}(d)]. The deviation of the ion velocity distribution from the initial Maxwellian is small [Figure \ref{fig:vphase_space_case2}(b)], but is required to reach the Vlasov equilibrium.

\begin{figure}[tphb]
    \centering
    \includegraphics[width=0.6\textwidth]{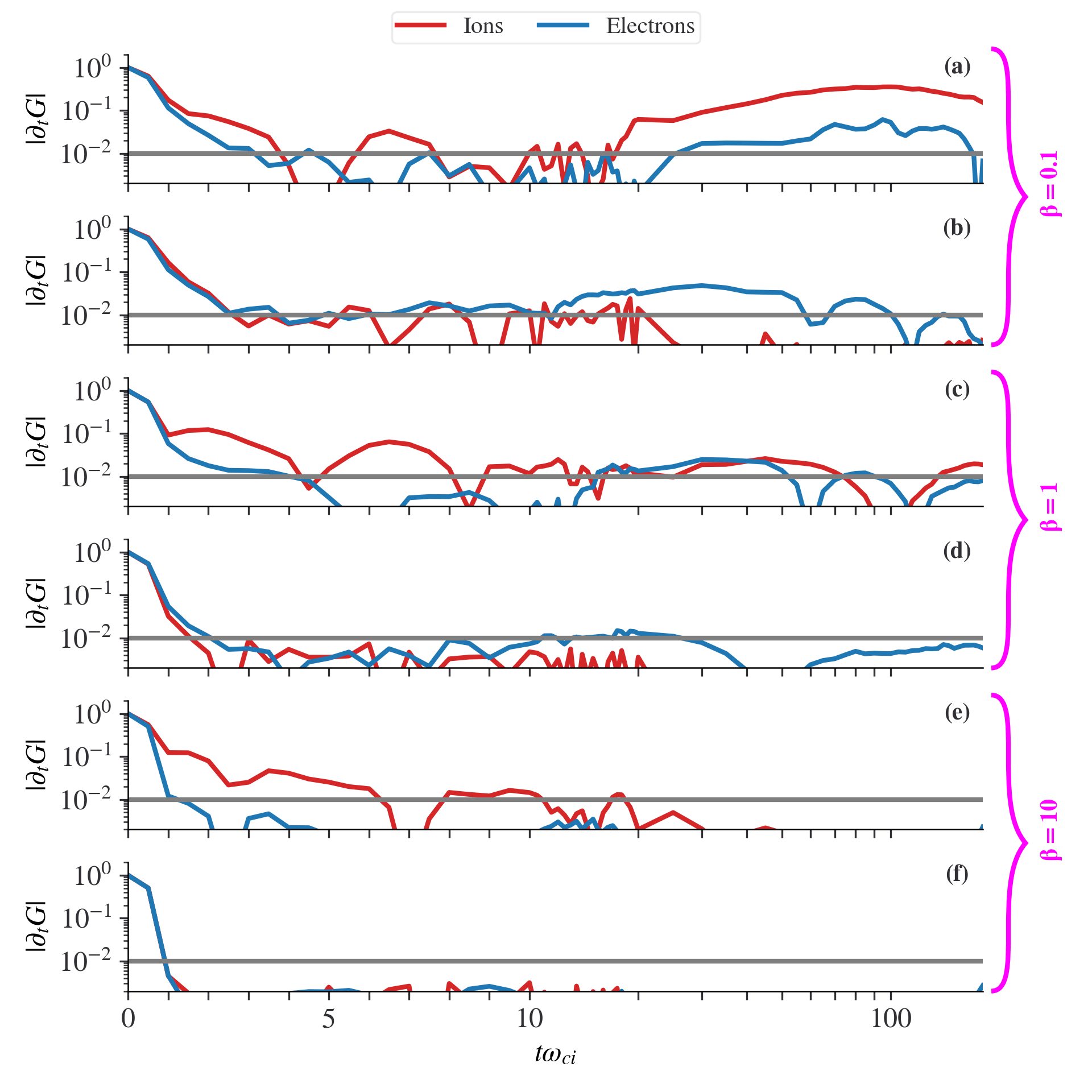}
    \caption{Evolution of force-free current sheets toward kinetic equilibrium with different plasma betas. The $x$-axis has two scales, a linear scale in $0 \leq t \omega_{ci}^{-1} \leq 10$ and a log scale in $10 \leq t \omega_{ci}^{-1} \leq 180$. The $y$ axis denotes $\vert \partial G / \partial t \vert$. The rows from top to bottom represent Runs 1A, 1B, 2A, 2B, 3A, and 3B, respectively. The grey line $\vert \partial G / \partial t \vert = 0.01$ in each panel is drawn as a reference.}
    \label{fig:reaching_kequibrium}
\end{figure}

To quantify the convergence of the system toward kinetic equilibrium, we define the metric
\begin{equation}
    G_s(t) = \int_{-\infty}^{\infty} \mathrm{d} v_\parallel \left\langle \left(g_s(x, v_\parallel, t) - g_s(x, v_\parallel, 0)\right)^2 \right\rangle_x ,
\end{equation}
where $\langle \cdot \rangle_x$ denotes the average over spatial coordinate $x$. This metric is chosen in such a way that the non-Maxwellian features of $g_s(x, v_\parallel, t)$ are properly accounted for. During the evolution of the system, it is expected that $G_s(t)$ experiences significant changes at the beginning of the simulation and eventually reaches a steady state, if a kinetic equilibrium is established. Figure \ref{fig:reaching_kequibrium} shows the absolute values of the time derivative $\vert \partial G_s /\partial t \vert$ for both ion- and electron-dominated current sheets with different plasma betas. All runs end up with $\vert \partial G_s /\partial t \vert \lesssim 0.01$, except for Run 1A (i.e., the ion-dominated current sheet with $\beta = 0.1$). For the same plasma beta, electron-dominated current sheets reach the kinetic equilibrium faster than ion-dominated current sheets. In the latter case, the system simply takes more time to redistribute the currents between unmagnetized ions and magnetized electrons. For both types (ion- and electron-dominated) of current sheets, those with higher plasma $\beta$ evolve toward kinetic equilibrium faster, because the current sheets with higher plasma $\beta$ can support larger transient electric fields to redistribute particles more rapidly in phase space toward the equilibrium.

At present, we cannot simulate further in time the relaxation of the low-$\beta$, ion-dominated current sheet in Run 1A toward the final kinetic equilibrium, because the current sheet becomes unstable at a later time (most likely due to boundary conditions). However, the progression of the relaxation for different beta values and the behavior of Run 1A up to $t \cdot \omega_{ci} \sim 10$ suggests that it also conforms to the explanatory model presented herein. The distribution functions at that time provide a sufficiently good representation of the Vlasov equilibrium to use in future modeling.

\section{Conclusion and discussion} \label{sec:conclusion}
In summary, we demonstrate that kinetic equilibria of 2D force-free current sheets exist for different proportions of ion and electron currents in various plasma betas. When initial currents are carried purely by ions, field-aligned electron currents are developed by transient parallel electric fields, while perpendicular electrostatic fields are generated due to unmagnetized ions and magnetized electrons. When initial currents are carried exclusively by electrons, the macroscopic state of the system remains unchanged from its initial state, described by the MHD model. In both scenarios, the electron and ion distribution functions at the late equilibrium states show systematic deviations from the initial drifting Maxwellians. These deviations occur in order to satisfy the time-stationary Vlasov equation.

\blue{We should mention that the proposed equilibrium must be constructed as a solution of the stationary Vlasov-Maxwell equations, but there is an infinite number of such solutions \citep{Grad61}. For each system, a specific solution should be determined by the boundary conditions and via relaxation (nonstationary) process. Therefore, our results show the existence of such 2D force-free current sheets, but the obtained solution should not be considered as unique. As an example, in Appendix \ref{appendix}, we compare a broad class of theoretical solutions \citep{Harrison09:prl,neukirch2009detailed,Neukirch20:jpp} for 1D force-free current sheets of $B_z = 0$ with PIC simulations. We find that despite the existence of an apriori equilibrium set by an analytical solution for a choice of a vector potential, the simulation converges to a different equilibrium corresponding to a different vector potential, and there is no apparent control on the final solution, that is one of an infinite number of such solutions.}

The existence of a kinetic equilibrium for 2D force-free current sheets indicates that there is at least one hidden symmetry in the system, implying the existence of an additional integral of motion (i.e., an additional invariant). Our system has five dimensions $N_D = 5$ in the phase space (2D in the coordinate space and 3D in the velocity space). The two known invariants are the total energy $H$, and the $y$ component of the canonical momentum $P_y = m v_y + e A /c$. The degrees of freedom of the system are $N_f = N_D - N_I$, where $N_I$ is the number of invariants. The particle phase space densities at equilibrium are constructed as a function of such invariants of motion. To fully describe such kinetic equilibrium, which we now know it does exist, we must have $N_I > N_f$, or equivalently, $N_I > N_D / 2$. Thus, $N_I$ is at least $3$ (in our case) and there is at least one hidden symmetry. Starting from particle trajectory data in the equilibrium electromagnetic fields, future research on this topic could make use of machine learning models to help find the number of invariants or even discover the analytic formula of these invariants \citep[e.g.,][]{Liu&Tegmark21:ML,Liu&Tegmark22:Poinkare2}.

\section{Acknowledgments}
This work was supported by NASA grants 80NSSC20K1788, 80NSSC22K0752, 80NSSC22K1634, and NAS5-02099. We acknowledge high-performance computing support from Cheyenne (doi:10.5065/D6RX99HX) provided by NCAR's Computational and Information Systems Laboratory, sponsored by the National Science Foundation \citep{cheyenne}.

%






\appendix

\section{\blue{Comparison between the theoretical solution and the PIC simulation for 1D force-free current sheets}\label{appendix}}
\blue{A broad class of theoretical solutions \citep{Harrison09:prl,neukirch2009detailed,Neukirch20:jpp} has been obtained for the stationary Vlasov-Maxwell equations of 1D force-free current sheets with $B_z=0$, i.e., for 1D force-free tangential discontinuities (also called 1D force-free Harris current sheet after \cite{Harris62} non-force-free solution). These solutions are quite useful, but not unique for the stationary Vlasov-Maxwell equations. Therefore, although there is almost no chance to obtain the same solutions via the relaxation method using PIC simulations (because of the infinite number of possible solutions), it may be informative and useful to demonstrate the applicability of our relaxation method by comparing its 1D results with these analytical solutions.}

\blue{The magnetic field of 1D force-free Harris current sheet is}
\begin{equation}\label{eq:mag-ffharris}
    \blue{\mathbf{B} = B_0 [\tanh(z/\lambda), 1/\cosh(z/\lambda), 0] .}
\end{equation}
\blue{The current density is}
\begin{equation}\label{eq:current-ffharris}
    \blue{\mathbf{j} = \frac{c B_0}{4 \pi \lambda} [\tanh(z/\lambda)/\cosh(z/\lambda), 1/\cosh^2(z/\lambda), 0] .}
\end{equation}
\blue{The vector potential in Coulomb gauge is given by}
\begin{equation}
    \blue{\mathbf{A} = B_0 \lambda [2 \arctan(e^{z/\lambda}), -\ln(\cosh(z/\lambda)), 0] .}
\end{equation}
\blue{The theoretical distribution function \citep{neukirch2009detailed} reads}
\begin{equation}\label{eq:neukirch2009-vdist}
    \blue{f_s =  \frac{n_{0s}}{\left(\sqrt{2 \pi} v_{th, s}\right)^3} \exp\left(-\beta_s H_s\right) \left[\underbrace{\exp\left(\beta_s u_{ys} p_{ys}\right)}_{\mathrm{Component\, 1}} + \underbrace{a_s \cos\left(\beta_s u_{xs} p_{xs}\right) + b_s}_{\mathrm{Component\, 2}}\right] ,}
\end{equation}
\blue{where $H_s$ is the Hamiltonian, and $\mathbf{p}_{s} = m_s \mathbf{v} + \frac{q_s}{c} \mathbf{A}$ is the canonical momentum. The subscript $s = \{i, e\}$ indexes ions and electrons. The constant parameters include the density $n_{0s}$, the inverse of temperature $\beta_s$, the thermal velocity $v_{th, s} = 1/\sqrt{\beta_s m_s}$, the characteristic drift velocities $u_{xs}$ and $u_{ys}$, and the dimensionless numbers $a_s$ and $b_s$. The distribution function consists of two components: Component $1$ is the familiar term carrying current in the $y$ direction as in the Harris current sheet; Component $2$ is the new term carrying current in the $x$ direction for the force-free Harris current sheet. There are constraints on $a_s$, $b_s$, $u_{xs}$, $u_{ys}$ and $v_{th,s}$ to ensure both the positivity of $f_s$ and a single maximum of $f_s$ in $v_x$ and $v_y$ \citep[to avoid possible microinstabilities in velocity space;][]{neukirch2009detailed}. In addition, the quasi-neutrality condition relates the ion parameters with the electron parameters. We summarize the calculation of these parameters from the macroscopic parameters $B_0$ and $\lambda$:}
\blue{
\begin{eqnarray}
    & \frac{2 c}{B_0 \lambda} = - e \beta_e u_{ye} = e \beta_i u_{yi} , \\
    & \frac{2 c}{B_0 \lambda} = e \beta_e u_{xe} = e \beta_i u_{xi} , \\
    & a = \frac{1}{2} = a_e \exp\left[-\frac{\beta_e m_e (u_{xe}^2 + u_{ye}^2)}{2}\right] = a_i \exp\left[-\frac{\beta_i m_i (u_{xi}^2 + u_{yi}^2)}{2}\right] , \\
    & b = b_e \exp\left(-\frac{\beta_e m_e u_{ye}^2}{2}\right) = b_i \exp\left(-\frac{\beta_i m_i u_{yi}^2}{2}\right) , \\
    & n_0 = n_{0e} \exp\left(\frac{\beta_e m_e u_{ye}^2}{2}\right) = n_{0i} \exp\left(\frac{\beta_i m_i u_{yi}^2}{2}\right) , \\
    & \frac{B_0^2}{8 \pi} = n_0 \left(\frac{1}{\beta_e} + \frac{1}{\beta_i}\right) .
\end{eqnarray}
}
\blue{In this study, we choose $m_i / m_e =100$, $\lambda = 2 d_i$, $T_i = 5 T_e = \frac{5}{12} m_i v_{\mathrm{A}}^2$ (equivalently, $\beta_i = 0.2 \beta_e = 2.4 (m_i v_{\mathrm{A}}^2)^{-1}$), and $b = 1$ (ensuring $f_s > 0$, as well as a single maximum of $f_s$ in $v_x$ and $v_y$). Then the other parameters are determined as $u_{yi} = u_{xi} = \frac{5}{12} v_{\mathrm{A}}$, $u_{ye} = -u_{xe} = -\frac{1}{12} v_{\mathrm{A}}$, $a_e = 0.50$, $a_i = 0.76$, $b_e = 1.0$, $b_i = 1.2$, $n_{0e} = n_0$, and $n_{0i} = 0.81 n_0$.}

\blue{In the PIC simulations, for the initial conditions we initialized a two-component Maxwellian that gives the same density, flow velocity, and temperature as that in Equation \eqref{eq:neukirch2009-vdist}:}
\begin{equation}\label{eq:initial-Maxwellian}
    \blue{f_{Ms} = \frac{n_{s1}}{(\sqrt{2 \pi} v_{th,s})^3} \exp\left[-\beta_s m_s \left(v_x^2 + (v_y - v_{ds1})^2 + v_z^2\right)\right] + \frac{n_{s2}}{(\sqrt{2 \pi} v_{th,s})^3} \exp\left[-\beta_s m_s \left((v_x - v_{ds2})^2 + v_y^2 + v_z^2\right)\right] ,}
\end{equation}
\blue{with
\begin{eqnarray}
    & n_{s1} &= n_0 / \cosh^{2}\left(\frac{z}{L}\right) , \\
    & n_{s2} &= n_0 \left[\frac{1}{2} + b - \frac{1}{\cosh^{2}\left(\frac{z}{L}\right)}\right] , \\
    & v_{ds1} &= u_{ys} , \\
    & v_{ds2} &= u_{ys} \frac{\sinh\left(\frac{z}{L}\right)}{\left(\frac{1}{2} + b\right) \cosh^2\left(\frac{z}{L}\right) - 1}  .
\end{eqnarray}
The boundary condition for both fields and particles is periodic in the $x$ direction, whereas the boundary condition in the $z$ direction is the same as that described in Section \ref{sec:setup}. With the magnetic field and distribution function given by Equations \eqref{eq:mag-ffharris} and \eqref{eq:initial-Maxwellian}, respectively, we let the system evolve until an equilibrium state is reached.}

\begin{figure}
    \centering
    \includegraphics[width=0.8\textwidth]{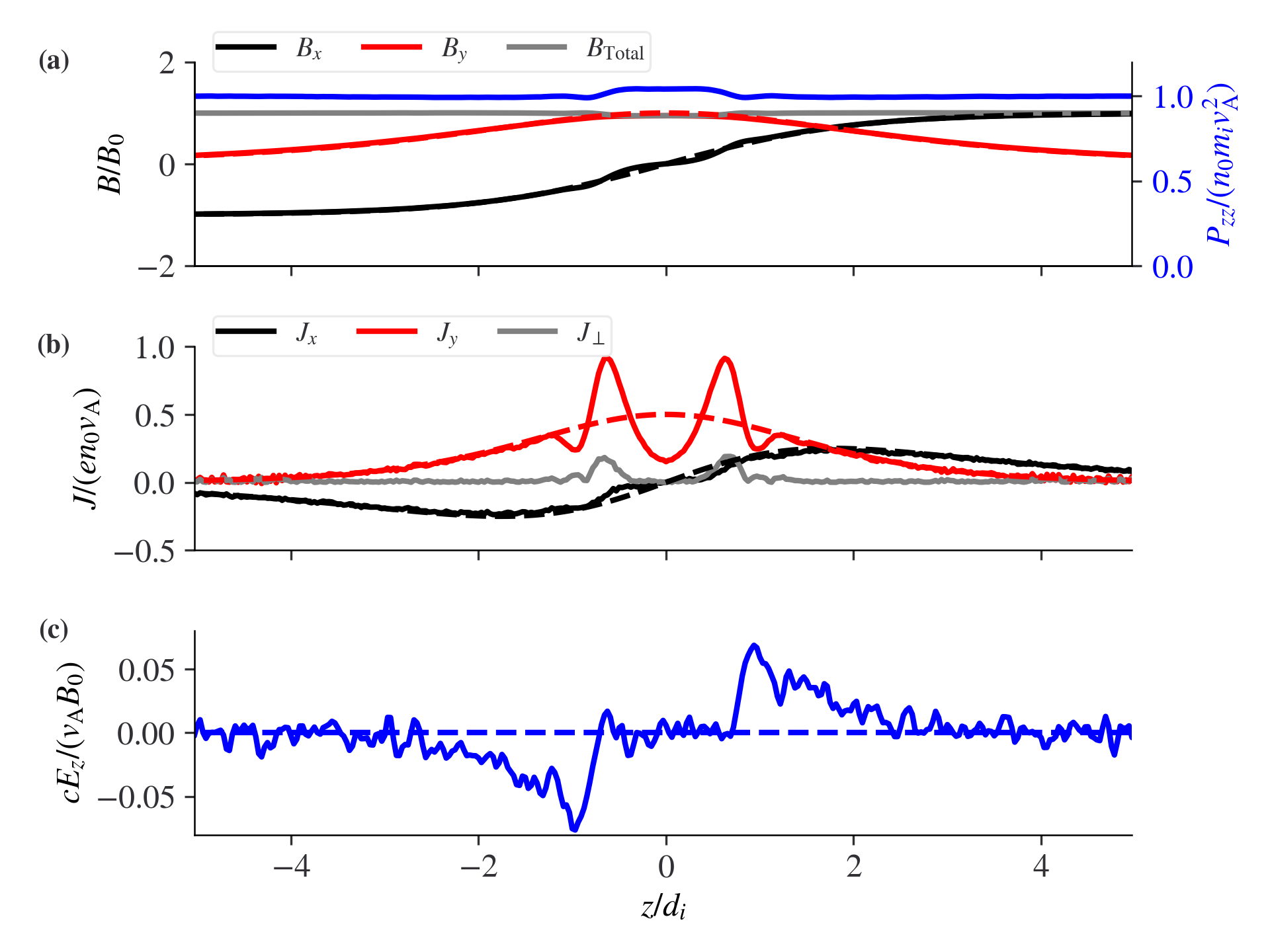}
    \caption{\blue{The final state of the system at $t = 1000\, \omega_{ci}^{-1}$. (a) Magnetic field components $B_x$, $B_y$, and their total strength $B_{\mathrm{Total}} = \sqrt{B_x^2 + B_y^2}$. The plasma pressure $P_{zz}$ is shown in blue. (b) Current density components $J_x$, $J_y$, and their projection $J_\perp$ perpendicular to the magnetic field. (c) Polarized electric field $E_z$. In each panel, the dashed lines represent the solution of the force-free Harris current sheet given by Equations \eqref{eq:mag-ffharris}, \eqref{eq:current-ffharris}, and $E_z = 0$.}}
    \label{fig:line-flds}
\end{figure}

\blue{Figure \ref{fig:line-flds} shows the macrostate of the system at the final equilibrium ($t = 1000\, \omega_{ci}^{-1}$). The system is in a force-free equilibrium as evidence by $P_{zz} \approx \mathrm{constant}$ and $J_\perp \approx 0$. Compared with the initial state, the current density (mainly $J_y$) is bifurcated, possibly due to the orbit class transitions \citep{Yoon21:NatCom,yoon2023equilibrium} in the relaxation process of the current sheet \citep[see discussion of such biffurcated current sheet models in, e.g.,][]{CL05,Sitnov06}. A polarized electric field points from the equator ($z = 0$) to higher latitudes, which is similar to the force-free Lembege-Pellat current sheet [Figure \ref{fig:efield}].}

\begin{figure}
    \centering
    \includegraphics[width=\textwidth]{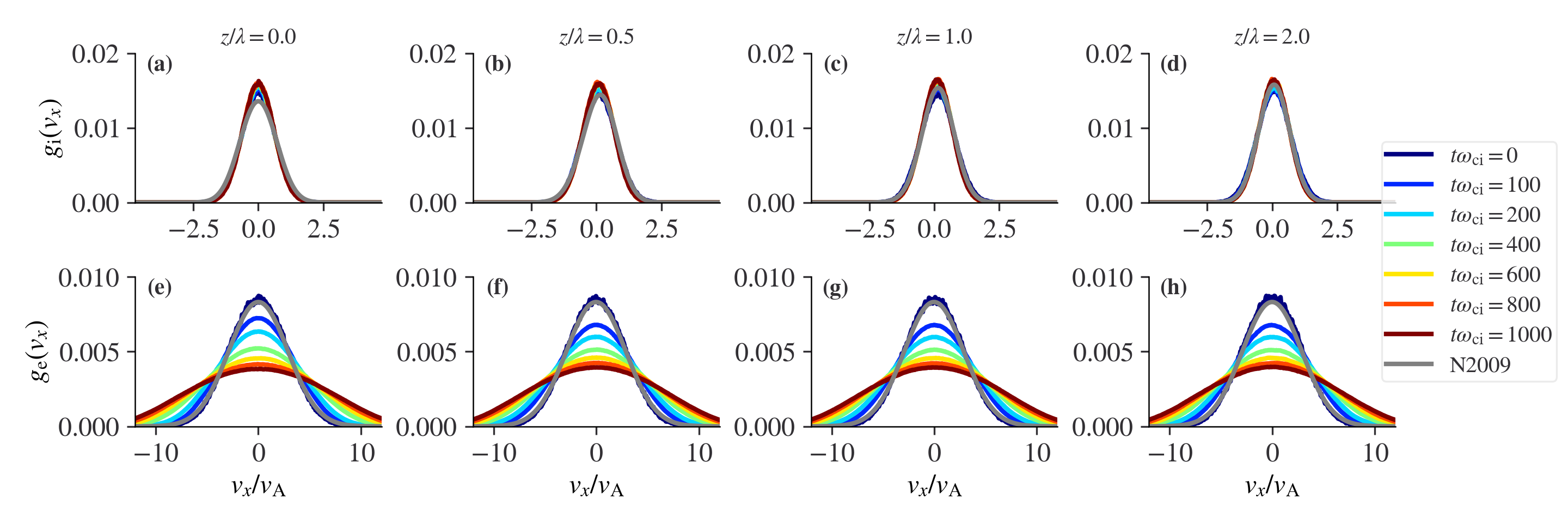}
    \caption{\blue{Time evolution of the reduced particle distribution function $g_s(v_x)$ ($s = i,\, e$). (a)-(d) The ion reduced distribution functions at $z / \lambda = 0, 0.5, 1, 2$. (e)-(h) The electron reduced distribution functions at $z / \lambda = 0, 0.5, 1, 2$. The distribution functions in each panel are shown from $t=0$ to $t=1000\, \omega_{ci}^{-1}$, colored-coded from blue to red. The grey curve in each panel labeled ``N2009'' represents the reduced distribution from Equation \eqref{eq:neukirch2009-vdist} \citep{neukirch2009detailed}.}}
    \label{fig:phase-space-z-vx}
\end{figure}

\blue{Figure \ref{fig:phase-space-z-vx} shows the evolution of the reduced distribution function $g_s(v_x) = \int_{-\infty}^{\infty}\int_{-\infty}^{\infty}\mathrm{d}v_y\mathrm{d}v_z f_s(v_x, v_y, v_z)$ at four representative $z$ locations. The ion and electron distribution functions asymptotically converge to an equilibrium state, which is different from Equation \eqref{eq:neukirch2009-vdist} \citep[][referred to as N2009 in Figure \ref{fig:phase-space-z-vx}]{neukirch2009detailed}. In particular, the equilibrium electron distribution has a substantially larger thermal velocity than the initial one, whereas the equilibrium ion distribution has a similar thermal velocity as the initial one. Because the solution is not unique for the stationary Vlasov-Maxwell equations of 1D force-free current sheets \citep{Grad61,Channell76}, the final equilibrium distribution function is not guaranteed to be the same as Equation \eqref{eq:neukirch2009-vdist}.}

\bibliographystyle{aasjournal}



\end{document}